\documentclass[preprint,showpacs,superscriptaddress,
footinbib,citeautoscript,floatfix]{revtex4-1}
\usepackage{graphicx}

\usepackage[utf8]{inputenc}
\usepackage{csquotes}
\usepackage[american]{babel}
\usepackage[T1]{fontenc}
\usepackage{enumerate}
\usepackage{mdwlist}
\usepackage[activate=normal]{pdfcprot}
\usepackage{bbding}
\usepackage{color}
\usepackage{amsmath}
\usepackage{eqnarray,amsmath}
\usepackage{multirow}
\usepackage[titletoc,title]{appendix}
\usepackage{amssymb}
\usepackage{amsmath}
\usepackage{amsfonts}
\usepackage{mathrsfs}
\usepackage[titletoc,title]{appendix}
\usepackage{empheq}

\usepackage{bm}
\usepackage{dcolumn}
\usepackage{color}
\usepackage[normalem]{ulem}
\usepackage{bbm}
\usepackage[colorlinks=true,citecolor=blue]{hyperref}
\hypersetup{colorlinks=true,citecolor=blue,linkcolor=red
	,urlcolor=blue}

\newcommand* {\bra}[1]{\ensuremath{\langle {#1} |}}
\newcommand* {\ket}[1]{\ensuremath{| {#1} \rangle}}
\newcommand* {\braket}[1]{\ensuremath{\langle {#1} \rangle}}

\newcommand{\beq}{\begin{equation}}
	\newcommand{\eeq}{\end{equation}}
\newcommand{\beqa}{\begin{eqnarray}}
	\newcommand{\eeqa}{\end{eqnarray}}
\DeclareMathOperator{\sech}{sech}

\begin{document}
	\title{Probing domain wall dynamics in magnetic Weyl semimetals via the non-linear anomalous Hall effect}
	\author{Shiva Heidari}
	\affiliation{School of Physics, Institute for Research in Fundamental Sciences, IPM, Tehran, 19395-5531, Iran}
	\author{Reza Asgari}
	\email{r.asgari@unsw.edu.au, asgari@ipm.ir}
	\affiliation{School of Physics, Institute for Research in Fundamental Sciences, IPM, Tehran, 19395-5531, Iran}
	\affiliation{School  of  Physics,  University  of  New  South  Wales,  Kensington,  NSW  2052,  Australia}
	\author{Dimitrie Culcer}
	\affiliation{School  of  Physics,  University  of  New  South  Wales,  Kensington,  NSW  2052,  Australia}
	\begin{abstract}
		The magnetic textures of Weyl semimetals are embedded into their topological structure and interact dynamically with it. Here, we examine electric field-induced structural phase transitions in domain walls mediated by the spin transfer torque, and their footprint in charge transport. Remarkably, domain wall dynamics lead to a transient, local, non-linear anomalous Hall effect and non-linear anomalous drift current, which serve as direct probes of the magnetization dynamics and of the domain wall location. We discuss experimental observation in state-of-the-art samples. 
	\end{abstract}
	\maketitle

	%\section{Introduction}  
	
	% Put the emphasis squarely on the non-linear AHE. 
	% Non-linear responses are very restrictive. They are used for optics, DC, and we show they can also reveal things we cannot see directly. Revealing the dynamics of complex systems. 
	% We need the symmetry reason why this transient NLAHE exists. There is always a gradient there in M(r). Which part of the response actually breaks inversion? 
	% First we introduce the DWs. We determine the DW motion and the associated NLAH current density, as well as the Hall voltage due to the non-linear Hall effect. We discuss experimental observation. 
	% We need to get the big figure on page 2. It is compiling very slowly.

	% There is a general structure of the 4-th rank tensor. 
	% Brataas 1503.01899.
	% Time scales: GHz for M, ps for e-.
	% Disorder in the general case.
	% DWM is dissipative dynamics. 
	% AHE & damping torque. 
	% S(r) looks very much like SHE. Chiral charge density.
	
	% NL helps us image/track things moving through inhomogeneous spatial distributions.
	% Does anything happen when the frequency of the electric field matches the time-scale over which the magnetisation changes?

	\section{Introduction}
	
	The interplay of topology and magnetism in condensed matter systems has spawned a series of fundamental scientific and technological research directions, such as magnetic topological insulators \citep{Tokura_2019,Wang_2021,Yao_2021}, magnetic Weyl semimetals (WSMs) \citep{Liu_2019,Morali_2019} and topological spintronics \citep{Fan_2016}. Coupling magnetism with topological phases of matter has led to the quantum anomalous Hall effect \citep{Yu_2010, PhysRevLett.106.166802, Chang167, Checkelsky_2012}, giant magnetoresistance \citep{PhysRevB.39.4828,PhysRevLett.61.2472}, current-induced spin manipulation \citep{Meier_2007,Kl_ui_2005,Vanhaverbeke_2008}, universal magneto-optical response \citep{PhysRevLett.105.057401,Okada_2016}, interface-induced magnetic phenomena \citep{RevModPhys.89.025006}, and electrical tuning of magnetizations at the interface between a topological insulator and a ferromagnetic insulator, with energy-saving applications in electronics \citep{PhysRevB.81.121401,PhysRevLett.104.146802,PhysRevB.82.161401,PhysRevLett.109.237203,PhysRevLett.108.187201,PhysRevB.89.024413,PhysRevB.90.041412,PhysRevB.92.085416,Kim_2019}. In particular, the detection and manipulation of topological defects, such as domain walls, using nanosecond spin-polarized current pulses provides a platform for storing and preserving information, for example using racetrack magnetic memory \citep{Parkin_2008} and logic gates \citep{Zhang_2015}.
	
	Domain wall (DW) dynamics can be driven by the electrically-induced spin-transfer torque (STT) \citep{RALPH20081190,Brataas_2012,PhysRevLett.92.086601,PhysRevB.54.9353,10.21468/SciPostPhys.10.5.102}, which also controls magnetization precession  \citep{PhysRevLett.80.4281,PhysRevLett.92.027201,Kiselev_2003,Krivorotov_2005,Myers_1999,PhysRevLett.84.3149,PhysRevB.54.9353,Slonczewski_1996}, as well as by the spin-orbit torque \citep{Spaldin_2019}. Interest in spin manipulation in topological materials has motivated an increasing number of empirical studies on DW dynamics in magnetic WSMs \citep{Destraz_2020,Suzuki_2019,Lee_2022,Howlader_2020,xu2021directional,sun2021mapping}, which have received considerable attention in recent years \citep{Liu_2018, Wang_2018, Sakai_2018, Belopolski_2019, Li_2020, Kuroda_2017, Hirschberger_2016, Borisenko_2019, PhysRevMaterials.4.024202, PhysRevX.9.041061, Guin_2019, PhysRevLett.123.187201, PhysRevLett.126.236601}.

	In magnetic WSMs the magnetic texture is inherent to the topological structure and exhibits a dynamical interplay with it. The momentum space locations of the Weyl nodes $\bm{k}_w$ are determined by the magnitude and direction of the magnetization, therefore any topological defects, such as domain walls, can move the nodes in momentum space, causing them to exhibit a space and time dependence of the form $\hat{k}_w=\hat{M}(r,t)$. The vital question in this context is understanding inhomogeneous current responses stemming from DW dynamics, which can detect magnetic texture dynamics in regions otherwise inaccessible. In this regard, we demonstrate that DW dynamics in magnetic WSMs lead to a transient, local, non-linear anomalous Hall effect (NAHE) and non-linear anomalous drift current (ADC). Such non-linear responses are the only non-linear signals of the system and are enabled by inversion symmetry breaking by the inhomogeneous magnetization.
 
	\begin{figure}[t]
		\centering
		\includegraphics[width=0.5\textwidth]{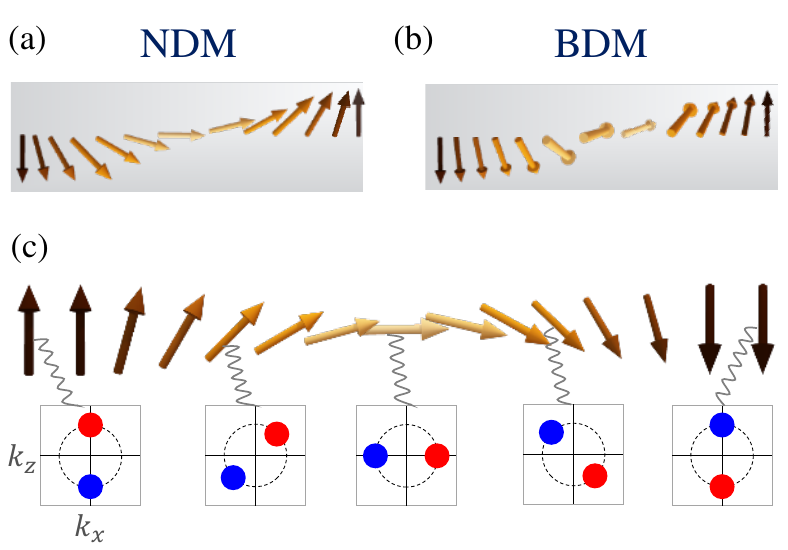} 
		\caption{(a) N\'{e}el domain wall (NDW) and (b) Bloch domain wall (BDW) in the static condition. (c): The coupling between magnetic texture in real space and Weyl nodes in momentum space. The red and blue circles represent the Weyl node positions with $\chi=+1$ and $\chi=-1$, respectively.} \label{fig_1}
	\end{figure}
	
	Whereas we focus here on magnetic WSMs, the physics we describe is general and the non-linear anomalous Hall effect stemming from domain wall motion has a straightforward dynamical explanation. To begin with, in an infinitesimal time interval $\delta t$ an electron spin at ${\bm r}$ is displaced to ${\bm r} + \delta {\bm r}$. The magnetization at the new spatial location, ${\bm M}({\bm r} + \delta {\bm r})$, has a different orientation from that at the original location, ${\bm M}({\bm r})$. Hence ${\bm M}({\bm r} + \delta {\bm r})$ exerts a torque on the electron spin and rotates it by a small amount. Since the spin is coupled to the wave vector, the spin rotation results in a small rotation of the wave vector. If we focus, for simplicity, on an initial wave vector parallel to the applied electric field, the rotation will result in a small transverse component, in other words a Hall effect. Thus far this is simply the linear Hall effect: if only the electron is in motion a trivial linear Hall effect results. However, in this case the magnetization is also a function of time. Thus, in a time $\delta t$, while the electron travels from ${\bm s}({\bm r})$ to ${\bm r} + \delta {\bm r}$, ${\bm M}({\bm r} + \delta {\bm r})$ has evolved in time, and the change in the magnetization is proportional to its gradient, and in turn to the electric field. This extra change in ${\bm M}$ induces an additional contribution to the Hall effect proportional to the square of the electric field: one factor for the electron acceleration and one factor for the change in magnetization. 
	\section{Theory and Model}
	The low-energy Hamiltonian describing three-dimensional (3D) WSMs coupled with local magnetic moments through the exchange interaction ${\cal J}$ is given by
	\begin{equation} \label{Hmodel}
		H_0 = \hbar v_{\rm F} \tau_z \otimes {\bm \sigma} \cdot {\bm k} - {\cal J} S \mathbbm{1} \otimes {\bm \sigma} \cdot \hat{\bm M}(r,t),
	\end{equation}
	where $\tau_z$ is acting on the chirality with eigenvalues $\chi=\pm 1$ and $\sigma_i$ are the Pauli matrices and $\bm{M}=S \hat{M}$ is the 3D magnetic texture vector with constant magnitude $S$. The $k$-space coordinate of the Weyl nodes in the ferromagnetic domains with $\bm{M}=\pm S \hat{z}$ is given by $\bm{k}_w=\pm (0,0, \chi k_0)$, where $k_0=\frac{{\cal J}S}{\hbar v_{\rm F}}$. The presence of a spatially inhomogeneous magnetic texture, e.g., walls between ferromagnetic domains, only alters the Weyl node coordinate but never changes the Weyl node separation $2|k_0|$ [Fig. \ref{fig_1}]. The spatio-temporal dependency of $\bm{M}(x,t)$ should be smooth in comparison to the typical length and time scale of the system, i.e., long-range and low-frequency magnetic texture. In magnetic WSMs, magnetic textures are mapped onto chiral electromagnetic fields coupled with opposite signs to the Weyl nodes \citep{Araki_2019, Kurebayashi_2019, Tserkovnyak_2021, Liu_2013, Liang_2019}. Using the definition of the $U(1)$ axial-vector field $\bm{A}_5=\frac{{\cal J}S}{e v_{\rm F}}\hat{M}$ \citep{Hutasoit_2014} and the axial current $ \bm{J}_5=\sum\limits_{\chi} \chi J_\chi= -ev_{\rm F} \left \langle \bm{\sigma}\right \rangle$, the last term in ${H_0}$ can be written as $\bm{J}_5 \cdot \bm{A}_5$, which means that spin-density operator acts like an axial current density coupled to the magnetization $\bm{M}$, and axial perturbations such as domain walls host an axial charge density. 

	The localized axial magnetic field $\bm{B}_5$, which is intrinsic to the spatially inhomogeneous magnetic configuration exists when the curl of the magnetic orientation is non-zero, i.e., $\bm{B}_5=\frac{\hbar k_0}{e}  \nabla \times \hat{M}$ \citep{Araki_2019}. The main consequence of this feature leads to the local STT, $\bm{T}_e=k_0 \hat{M}\times \left \langle \bm{\sigma}\right \rangle$ in which the non-equilibrium spin polarized electrons $\left \langle \bm{\sigma}\right \rangle \propto k_0 \bm{B}_5 \times \bm{E}$ is induced by an external electric field \citep{Kurebayashi_2019,Araki_2019}. Similarly, the time-dependence of $\bm{A}_5$, gives rise to an axial electric field $\bm{E}_5 = -  d\bm{A}_5/dt$. Since the averaged pseudofields vanishes over the whole sample, the domain walls are arranged in such a way as to cancel each other out after summing over the entire sample. It is a general statement and does not depend on the domain wall type or domain (wall) width. We will demonstrate how the STT-induced $\bm{E}_5$ could contribute to the non-linear electronic responses, i.e., $J_y=\sigma_{yxx}E_x^2$, where $\sigma_{yxx} \propto B_5$. Therefore, non-linear signals in magnetic texture (DWs) must have a non-zero real space curvature. This originates from breaking the local mirror symmetry.

	%Using the \textit{Landau-Lifshitz-Gilbert} equation \citep{landau1935physik} in the presence of an electric field, the STT satisfies the following equation:
	
	%	\begin{equation} \label{STT}
		%	\dfrac{d\bm{M}}{dt}=\exp(\alpha \hat{\Theta}) \bm{T_e},
		%	\end{equation}
	%where $\hat{\Theta} \vec{\cal{O}}=\hat{M} \times \vec{\cal{O}}$, and the coefficient $\alpha$ is the Gilbert parameter  which is proportional to the rate of energy loss \citep{malozemoff2016magnetic}.
	%\begin{equation} \label{STT}
	%	\bm{T}_e=\dfrac{\sigma^{(H)}}{e \rho_s} \hat{M}\times[({\bf E}\cdot{\bf \nabla}) \hat{M} - {\nabla}(\hat{M}\cdot{\bf E})],
	%\end{equation}
	%	where $\sigma^{(\text{H})}_{\text{SC}}=-\frac{v_F \tau^2 e^3 \mu}{6\pi^2 \hbar^3} |B_5|$, with ${\bm B}_5 = {\bm \nabla}\times{\bm A}_5$, is valid for chemical potentials $\mu$ far above the first pseudo-Landau level (semiclassical limit) and $\sigma^{(\text{H})}_{\text{Q}}=\frac{e^2}{4\pi^2 \hbar^2} \frac{\mu}{ v_{\rm F}}$ implies the quantum limit when the Fermi level is near the crossing point \citep{Araki_2019}. In this paper we neglect the pseudo-Landau quantization and then use the semiclassical expression for $\sigma^{(\text{H})}$, and assume no external magnetic field is present.
	We focus on magnetic WSMs with inversion symmetry e.g. Co$_3$Sn$_2$S$_2$ \citep{Morali_2019}, HgCr$_2$Se$_4$ \citep{PhysRevLett.107.186806} and PrAlSi \citep{PhysRevB.102.085143,PhysRevB.104.014412}, which allows us to neglect the Dzyaloshinskii-Moriya interaction, as well as shape and dipolar anisotropy. Such terms may be strong in e.g. layered materials \citep{PhysRevLett.107.127205, PhysRevB.103.094433, PhysRevB.101.115106}, and may affect DW motion, but they do not change our fundamental argument that DW motion in WSM is accompanied by a non-linear electrical response, which can be used to probe the magnetic inhomogeneity.
	%We therefore study DW dynamics under the influence of ${\bm T}_e$ for the remainder of this paper. The STT $\bm{T}_e$  describes the background electronic contribution to the magnetic dynamics, arising from the exchange interaction between conduction electrons and magnetic moments.  
	
	\begin{figure*}[t]
		\includegraphics[width=\textwidth]{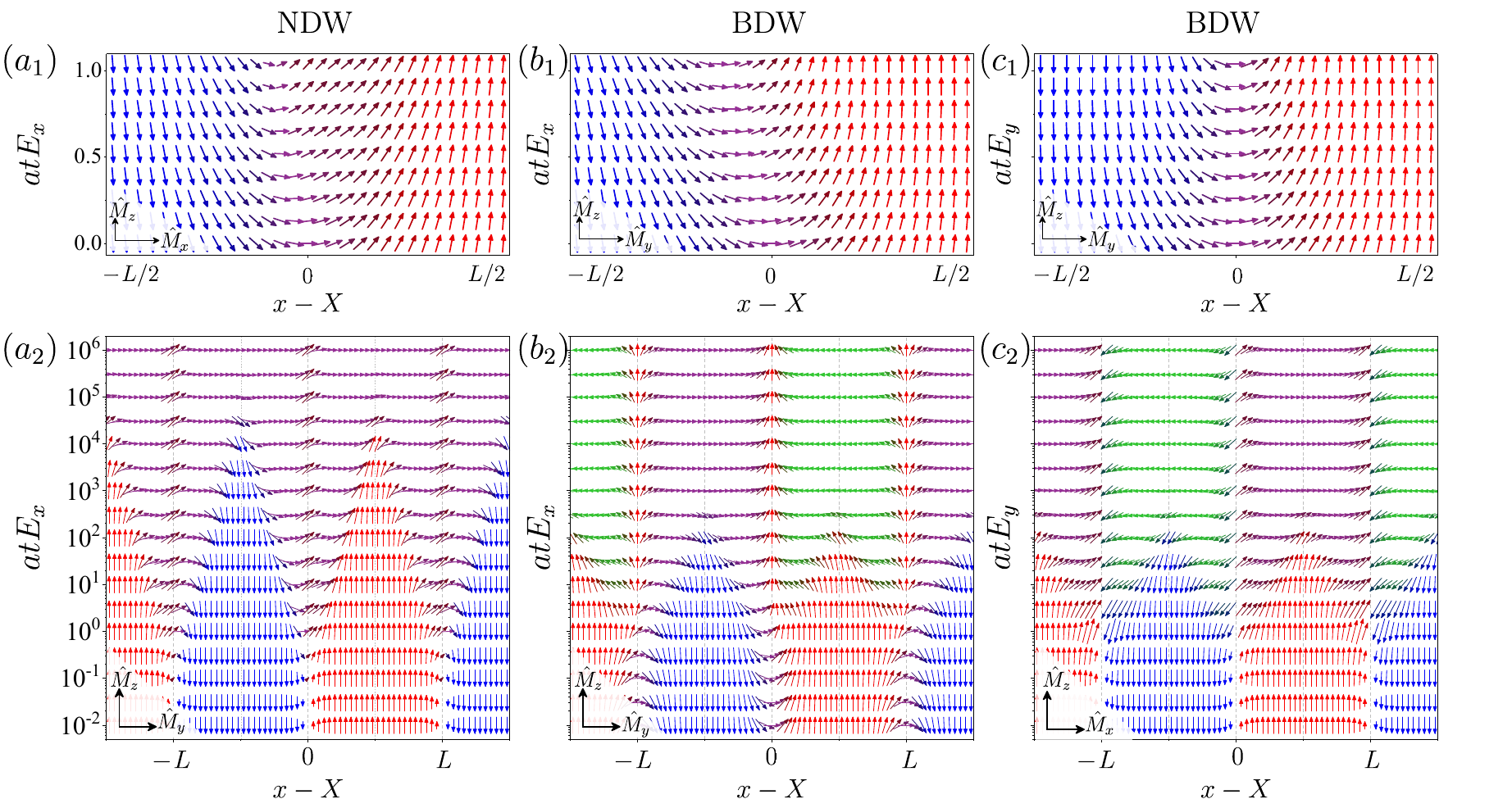}
		\caption{Electric-field-induced DW dynamics in WSMs. ($a_1$) and ($b_1$) represent the DW motion without structural deformation in response to $\bm{E}\parallel \hat{x}$ and $(c_1)$ shows that there is no rigid motion when $\bm{E}\parallel \hat{y}$. In the limit of $at|E| \geq 1$, the magnetic configuration shows a structural deformation of domain walls and in the higher limit when $at|E| \gg 1$, there will be a full magnetic transition of the domain texture. ($a_2$): A NDW between two ferromagnetic domains with $\pm \hat{M}_z$ disappears under the action of $E_x$ as the magnetization is rotated to $\hat{y}$. $(b_2(c_2))$: Applying an electric field along $x$-($y$-) to a BDW leads to a structural transition to the head-to-head (tail-to-tail) DW along $\hat{M}_y$($\hat{M}_x$). Here, X and L represent the DW center and ferromagnetic width, respectively.} \label{fig_2}
	\end{figure*}
	\subsection{Structural phase transition of magnetic texture}
	The electrically induced spin dynamics is described by the \textit{Landau-Lifshitz-Gilbert} equation \citep{landau1935physik}
	\begin{equation} \label{dyn1}
		\dfrac{d \hat{M}}{dt}= \gamma_0 {\bf B}_{eff} \times {\hat{M}}+\alpha \hat{M}\times\dfrac{d \hat{M}}{dt}+\bm{T}_e ,
	\end{equation}
	where $\gamma_0$ is the gyromagnetic ratio and ${\bf B}_{eff}$ is an effective magnetic field. The coefficient $\alpha$ is the \textit{Gilbert} or \textit{viscous damping parameter} which is proportional to the rate of energy loss \citep{malozemoff2016magnetic}. The torque $\bm{T}_e$ is a spin-transfer torque describing the background electronic contribution to the magnetic texture dynamics, arising from the exchange interaction between conduction electrons and magnetic moments. The torque $\bm{T}_e$ can be expressed in terms of the axial electric field $\bm{E}_5$, and Eq.~\ref{dyn1} is solved recursively as
		\begin{equation} 
				\hat{M}(x,t)\mid_{\text{New}}=f_E(x,t)^{-1}\bigg[\hat{M}(x,t)\mid_{\text{Old}}-\dfrac{e}{\hbar}\int_0^t \frac{\bm{E}_5(x,t^\prime)}{k_0}dt^\prime \bigg],
				\label{Mtt}
			\end{equation}
	 where $\bm{E}_5=\frac{\hbar}{e} k_0 (\bm{T}_e+\alpha \hat{M}\times \bm{T}_e)$ with the damping parameter $\alpha$ and the normalization factor is $f_E(x,t)=\sqrt{ \mid \hat{\bm M}-e/\hbar \int_0^t \frac{\bm{E}_5(x,t^\prime)}{k_0}dt^\prime\mid^2}$. 
	We define $a=(e \alpha \tau^2 v_F \mu |k_0|^2 )/(6 \pi^2 \rho_s \hbar^2 w)$ where $w$ is the DW width, and $\rho_s$ is the number of local magnetic moments per unit volume. 	We consider two orientations of the electric field: parallel and perpendicular to the hard axis of each DW, that is, $\bm{E} \parallel \hat{x}$ and $\bm{E} \parallel \hat{y}$. The DWs at $t=0$ are described by the vector $\hat {M}(x)=((1-\beta)\sech(\frac{x-X}{w}),\beta \sech(\frac{x-X}{w}) ,\tanh(\frac{x-X}{w}))$, where $\beta=0 (1)$ stands for N{\'e}el (Bloch) DW. The top panel of Fig. \ref{fig_2} represents the time evolution of DWs in the rigid motion regime, where $aEt<1$, i.e., dynamical motion without structural phase transition (SPT). In the case of BDW and $\bm{E}\parallel \hat{y}$, shown in Figs. \ref{fig_2}($c_1$) and $(c_2$), the axial electric field vanishes at $x=X$ and hence the spin moment right at the center freezes and then the DW does not undergo a rigid motion in space. The Walker breakdown regime \citep{Schryer_1974} starts when $aE t \geq 1$, leading to a SPT in both DWs and ferromagnetic domains. In particular, in Fig. \ref{fig_2}($a_2$) the NDW beyond the Walker breakdown regime disappears and the new magnetic structure near the center is given by $\alpha(y,1/\alpha,1-2y^2)/\sqrt{1+\alpha^2}$ where $y=x-X$. The BDW in Figs. \ref{fig_2}($b_2$) and ($c_2$), on the other hand, changes its structure to the head-to-head and tail-to-tail configurations, respectively. The steady-state condition for NDW and BDW, respectively, is established beyond $aE_xt \gtrsim 10^5$ and $aE_xt \gtrsim 10^3$ which means, interestingly, that the BDW approaches its steady-state much sooner than the NDW. The electric field along $\hat{y}$ does not induce any dynamics for NDW since the torque $\bm{T}_e|_{E_y}^{\text{NDW}}=0$.
	
	Therefore, the major characteristic of the STT in WSMs that causes structural phase transitions (SPT) in both the ferromagnetic domains and domain walls is that, depending on the direction of an external electric field, the NDWs can either vanish simultaneously or even be resistant to structural deformation. When the electric field is parallel to or perpendicular to the hard axis of the DW, the BDW also modifies its structure to the head-to-head or tail-to-tail configurations, respectively. 
	
	In general, we conclude that the system exhibits four different dynamical regimes: (i) rigid-motion, $atE<1$, where the DW moves without structural deformation with velocity  $v_D=\frac{|k_0| \sigma^{(H)}}{e \alpha \rho_s} \frac{c_1}{c_0} E_x$ \citep{Kurebayashi_2019} where $c_0= \int |\nabla_x \hat{M}|^2 \ dx$ and
	$c_1= \int (|\nabla_x \hat{M}|^2-|\nabla_x \hat{M}_x|^2)\  dx$, (ii) Walker-breakdown, $atE \geq 1$, where the DW stops moving and begins to undergo a structural transition (iii) ferromagnetic domain reconfiguration, $atE \gg 1$, and (iv) steady-state, $atE \ggg 1$.
\section{electrical signals of domain wall dynamics}
In previous sections, we have discussed the electric field-induced STT and how it is employed to change the magnetic structure. The STT, together with intrinsic $\bm{B}_5$ across the domain wall, is the main origin of the presence of $\bm{E}_5$ or magnetic texture dynamics. Such localized pseudo-electromagnetic fields like ordinary electromagnetic fields result in the electronic responses of intrinsic charge accumulation near the domain walls. 
Therefore, the temporal and spatial modulation of the ferromagnetic orders in magnetic texture induces anomalous electronic charge pumping from a macroscopic point of view \citep{PhysRevB.98.045302}. Having determined the domain wall dynamics in the presence of an external electric field together with axial fields, we are in a position to calculate the localized inter- and intra-band electronic currents using the quantum kinetic equation.

	Therefore, the spatially and temporally inhomogeneous magnetic configurations affect the dynamics of itinerant electrons, and are captured by the quantum kinetic equation \citep{Sekine_2017}
	\begin{eqnarray}\label{QKE1}
		\dfrac{\partial \braket{\rho_i}}{\partial t}+& \dfrac{i}{\hbar} [H_0,\braket{\rho_i}]+\dfrac{1}{2\hbar} \lbrace \dfrac{D {H}_0}{ D k} \cdot \nabla \braket{\rho_i} \rbrace + \kappa_i(\braket{\rho_i})\\ &={\cal D}_{M_t}(\braket{\rho_i})+{\cal D}_{M_r}(\braket{\rho_i})+{\cal D}_{E}(\braket{\rho_i}). \nonumber
	\end{eqnarray}
	Here $H_0$ is the band Hamiltonian given by Eq.~(\ref{Hmodel}), $\braket{\rho_i}$ is the density matrix of the i$^{th}$ node averaged over disorder, $\kappa(\braket{\rho_i})$ is the scattering integral, the electric-field-induced driving term is 
	${\cal D}_{E}(\braket{\rho_i})=e/\hbar \bm{E} \cdot \frac{D\braket{\rho_i}}{Dk}$ \citep{Culcer_2017}
	and the temporal and spatially inhomogeneous driving terms are given by [Appendix. \ref{appc}]
	\begin{align}\label{mt}
		{\cal D}_{M_t}(\braket{\rho_i}) & =\dfrac{-\chi {\cal J}S}{\hbar v_{\rm F}} \dot{\hat{M}} \cdot \dfrac{D\braket{\rho_i}}{Dk}, \\
		{\cal D}_{M_r}(\braket{\rho_i}) & =\dfrac{\chi {\cal J}S}{2 v_{\rm F} \hbar^2} \bigg\lbrace \dfrac{D H_0}{Dk}\times (\nabla \times \hat{M}) \cdot \dfrac{D\braket{\rho_i}}{D k} \bigg\rbrace . \label{mr}
	\end{align}
	The covariant derivative is defined as $\frac{D {\cal O}}{Dk}=\nabla_k {\cal O}-i[\bm{{\cal R}}_k,{\cal O}]$ where $\bm{{\cal R}}_k=\sum\limits_{a=x,y,z} {\cal R}_k^a e_a$ is the (vector) Berry connection with components $[{\cal R}_k^a]^{mn}=i\braket{u_k^m|\partial_{k_a}u^n_k}$. The electron-texture coupling terms are embedded in driving terms in Eq. (\ref{mt}), which can be used to derive ${\bm T}_e$.
	
	 The density matrix $\braket{\rho_{i}}$ can be decomposed into band-diagonal and band off-diagonal components, i.e., $\braket{\rho_{i}}^{nn^\prime}=\braket{\xi_{i}}^{nn^\prime} \delta_{nn^\prime}+(1-\delta_{nn^\prime})\braket{S_{i}}^{nn^\prime}$. The electronic current densities induced by DW motion originate from the intra- ($n \rightarrow n$) and inter-band ($n \rightarrow n^\prime$) excitations and can be obtained as $ \bm{J}_t=\text{Tr}[\bm{v}_k^{n} \braket{\xi_t}^{nn}_k+\bm{V}_k^{nn^\prime} \braket{S_t}_k^{n^\prime n}]=\bm{J}^{\text{d}}+\bm{J}^{\text{IB}}$ where $\bm{v}_k^n=\hbar^{-1}\nabla_k \epsilon_k^n$, $\bm{V}_k^{nn^\prime}=i \hbar^{-1}(\epsilon_k^n-\epsilon_k^{n^\prime}) {\cal R}_k^{nn^\prime}$, the operator trace Tr$=\sum_n \int [d\bm{k}]$  and subscript $t$ indicates it is induced by the STT. The intra-band contributions are contained in the diagonal part of the density matrix $\braket{\xi}$, while inter-band contributions are contained in the off-diagonal part $\braket{S}$, with band off-diagonal terms $\braket{S_t}_k^{nn^\prime} = \braket{S_E}_k^{nn^\prime}+\braket{S_{M_t}}_k^{nn^\prime}$ in response to an external electric field (labeled by subscript $E$) and magnetic texture dynamics (labeled by subscript $t$).
	
	Therefore, the non-linear current is found by taking the trace of the current operator with the density matrix, and takes the form [Appendix. \ref{appe}]
	\begin{equation} \label{interc}
		\bm{J}^{\text{IB}} = -\dfrac{e^2}{\hbar}|k_0|\dfrac{\hat{M}(x,t)\times \bm{E}}{f_E(x,t)}+\dfrac{e^3}{\hbar^2} \int_0^t \dfrac{\bm{E}_5(x,t^\prime) \times \bm{E}}{f_E(x,t)}dt^\prime,
	\end{equation}
	where $f_E(x,t)$ is a normalization factor. This expression denotes the AHE in WSMs in which the node distance along $\hat{M}_z$ is replaced by $\hat{M}(x,t)$ in the dynamical regime leading to the non-linear behavior in the anomalous Hall current ($E_5 \propto k_0^3 E$). We note that the first and second terms in Eq. (\ref{interc}) contain the linear and non-linear AHE proportional to $k_0 \ (={\cal J}S /\hbar v_f)$ and $k_0^3$, respectively. These two terms can be likely distinguished experimentally by imposing the mechanical force to reduce the Fermi velocity. The other non-linear terms are suppressed by inversion symmetry. 
		\begin{figure}[t]
		\includegraphics[width=0.5\textwidth]{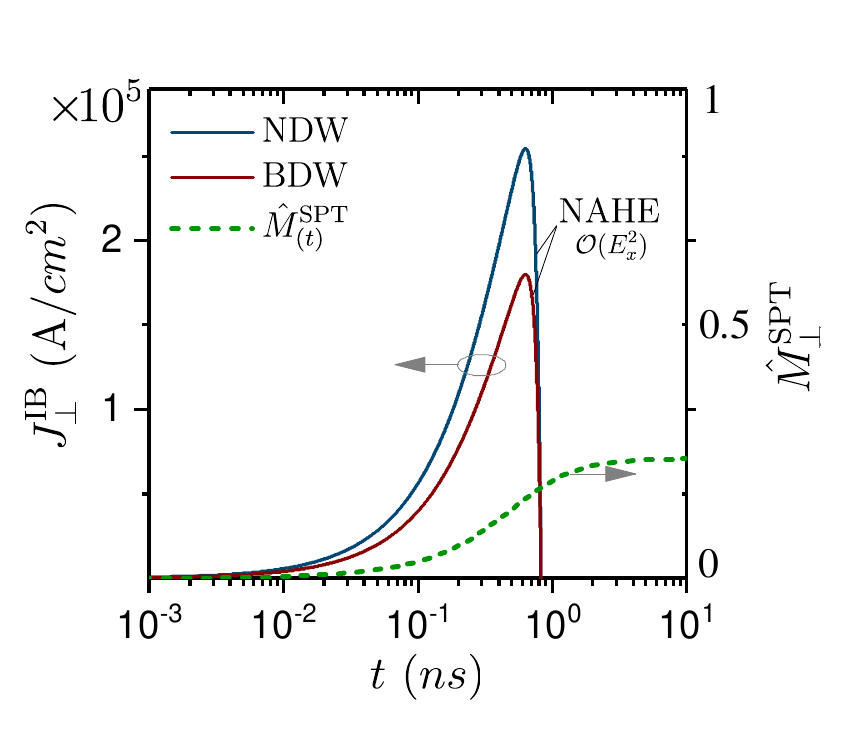}
		\caption{Non-linear anomalous Hall effect (NAHE) induced by the SPT of N{\'e}el domain wall (NDW) and Bloch domain wall (BDW). The non-linear response which measured at the center of DWs is represented as a function of time in a  fixed electric field $E=10^5$ V/m along the $x$-axis. The dashed green line shows the SPT dynamical regime of magnetic moment perpendicular to the hard axis of each DWs. The generation of $\hat{M}_\perp^{\text{SPT}}$ over time indicates the SPT in DWs. We set $w\approx 100$ nm, $\mu=100$ meV, $|2 k_0|=0.3$ \AA$^{-1}$.} \label{fig_3}
	\end{figure}

	Figure. \ref{fig_3} represents the $x=X$ value of the non-linear and transient AHE which signalizes the SPT in the DWs.
	The calculated transient peak of the NAHE is considerably stronger over an interval of a few nanoseconds. At $t=0$  and at the center of DW, the magnetic moment is along the hard axis ($\hat{x}$ for NDW and $\hat{y}$ for BDW), and as time evolves, the STT generates the magnetic component perpendicular to the hard axis  leading to the SPT in DWs.
	 Therefore, the SPT dynamical regime occurs when the magnetic component perpendicular to the hard axis, $\hat{M}_\perp^{\text{SPT}}$, is generated, and the new DW configuration is constructed. In the steady-state the non-linear signal vanishes, while the surviving linear response corresponds to the ordinary static AHE due to the equilibrium magnetic structure, i.e., $\bm{J} (aEt \ggg 1) \propto \hat{M}^{\text{New}} \times \bm{E}$. Existing proposals for the NAHE rely on the Berry curvature dipole, an entirely different mechanism active in non-magnetic and non-centrosymmetric materials \citep{Sodemann_2015, Battilomo_2019, Ma_2018, du2021perspective}. 
	
	To estimate the non-linear Hall 
	voltage, we use $E_y=j_x \rho_{yx}+j_y \rho_{yy}$ and the conventional relationship between the resistivity and conductivity tensors. After straightforward calculations and considering dominate conductivity elements, the second-order induced electric field along $\hat{y}$ is given by $E^{(2)}_y\approx\frac{\chi^{(2)}(t)}{\sigma_{yy}}E^{(2)}_x$ where $\chi^{(2)}(t)=\frac{e^3 \mu \tau^2 v_F}{6\pi^2\hbar^3\rho_s w}\frac{|k_0|^3 t}{f_E(0,t)}$ (extracted from Eq. (\ref{interc})) and $\sigma_{yy}=\frac{e^2 \mu \tau}{6\pi\hbar^3 v_F}$. Using the prior parameters for ${\bf J}^{\text{IB}}$, the induced-electric field is $E^{(2)}_y\approx7.15\times 10^6$ Vm$^{-1}$ which yields the non-linear Hall voltage $V^{(2)}_y=E^{(2)}_y d=0.71$ V for $d=w=100$ nm. This is straightforwardly measured.  
	%	\textcolor{red}{To estimate the $t=0$ value of the anomalous Hall current in Fig. \ref{fig3p}, we set the parameters of Co$_3$Sn$_2$S$_2$ as a magnetic WSM [Figure caption] 
		%	\begin{eqnarray}
			%	    & J_{t=0}=\dfrac{e^2}{\hbar} |k_0| \hat{M}\times \bm{E}= \nonumber \\ &\dfrac{e^2}{6.58\times 10^{-16} \text{eV.s}}\times 0.15 \times 10^4 [\text{1/m}] \times 10^5 [\text{V/m}])\nonumber \\
			%	    & =3.5 \times 10^5 \text{A/cm}^2 \nonumber
			%	\end{eqnarray}
		%	This is compatible with the result of \citep{Wang_2018}}
	%		\begin{figure}[t]
		%		\centering
		%		\includegraphics[scale=0.8]{fig_AHE.pdf} 
		%		\caption{adapted from \citep{Wang_2018}} \label{fig_AHE}
		%	\end{figure}.}
\begin{figure}[t]
	\includegraphics[width=0.5\textwidth]{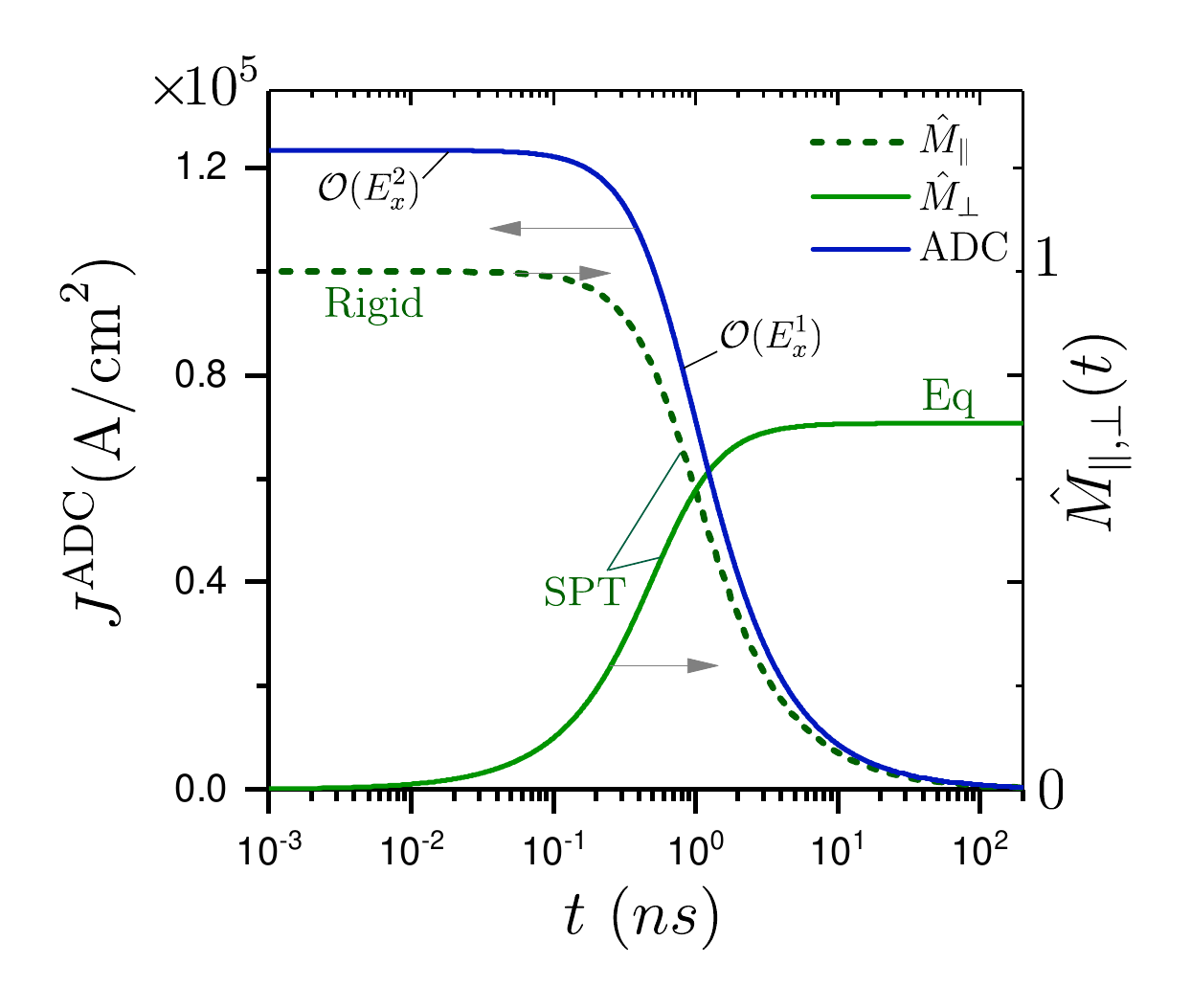}
	\caption{ The anomalous drift current (ADC) in response to the axial electric field $\bm E_5$. It inherits the non-linear response to $\bm E$ in the dynamical regime of DW and damps over time by approaching the steady-state. The dashed and solid lines represent the dynamical evolution of magnetic moment parallel and perpendicular to the hard axis of DW. The parameters chosen correspond to Fig. \ref{fig_3}.
	} \label{fig_4}
\end{figure}

Magnetization dynamics likewise leads to an intra-band drift current. The full expression is induced by magnetic dynamics $\dot{\hat{M}}$ is given by
\begin{equation} \label{drift}
\bm{J}_{\chi}^{\text{d}}=-e \chi |k_0| \sum_n \int [dk] \bm{v}_k^n \tau_k \nabla_k f_{0n} \cdot \dot{\hat{M}}.
\end{equation}
The drift current leads to the ADC in response to an axial electric field, i.e., $\bm{J}_\chi^{\text{d}}=\chi \sigma_\chi^{(\text{D})} \bm{E}_5$ with $\sigma_\chi^{(\text{D})}=\frac{-e \tau_d \mu_\chi^2}{3 \pi^2 \hbar^3 v_{\rm F}}$ where $\tau_d$ is the relaxation time for the drift motion \citep{Araki_2019,Araki_2018}. In order to estimate $\sigma_\chi^{(\text{D})}$, we set  $\mu_-=\mu_+=100$ meV, $v_{\rm F}=10^6$ m/s and $\tau_d=1$ ps, hence the drift conductivity for each node in a WSM is  $-2\times 10^5$ S/m. The total ADC after summing over two nodes is given by $\bm{J}^{d}=4\sigma^{(D)}  \frac{\delta \mu}{\mu} \bm{E}_5$ where $\delta \mu=|\mu_{+}-\mu_{-}|=(e^2\hbar \tau_d v_{\rm F}^3/\mu^2)\bm{E}_5 \cdot \bm{B}_5$. To estimate the average of axial magnetic field over a single DW, we take ${\cal J}S=1$ eV, $w=100$ nm, which gives $\bm{B}_5|_{\text{ave}} \simeq 2$ T (valid in semiclassical regime, i.e., $\mu \gg \hbar \tau^{-1}$ and $\mu \gg \epsilon_B$ where $\epsilon_B=v_{\rm F} \sqrt{e\hbar |B_5|}$), and $\bm{E}_5 \simeq 3 \times 10^4$ V/m, if $\delta \mu/\mu=5$\%.
The non-zero axial anomaly $\bm{E}_5 \cdot \bm{B}_5 \neq 0$, an imbalance in the number of left- and right-handed Weyl fermions, can exist for NDW by only applying an electric field along  $\hat{x}$, i.e., $\bm{E}_5 \cdot \bm{B}_5 \propto E_x \sech((x-X)/w)/f_E(x,t)$ which is a transient signal and only appears in the dynamical regime [Appendix. \ref{appf1}]. Therefore, the NDW dynamics can induce the non-linear ADC when $\bm E \parallel \hat{x}$.
It is also shown that the domain wall motion could induce such an axial anomaly in the presence of a magnetic field beyond the Walker Breakdown, which results in an oscillating current instead of a transient signal \citep{PhysRevB.102.241401,10.21468/SciPostPhys.10.5.102}. Figure \ref{fig_4} represents the non-linear ADC exists in a dynamical regime of NDW (rigid motion and SPT regime) and decays by approaching to the equilibrium. As a result, the observation of such unusual and transient responses can be employed as a potential candidate for confirming the existence of domain walls and their dynamical evolution in magnetic WSMs. For instance,
scanning probe-assisted techniques such as scanning tunneling microscopy and the detection of dynamical behavior of absorption coefficient as a transient signal at the infrared frequencies seems feasible to capture the local charge dynamics \citep{xu2020electronic,hsu2017electric,romming2013writing}. On the other hand, the motion of topological spin textures and domain walls can be captured by different techniques such as nonlinear magnetic resonance, magnetic force microscopy or magneto-optical measurements \citep{stepanova2021detection,jiang2017direct,kovacs2017mapping}. 
%	Furthermore, applying an electric field along $\hat{y}$ leads to a local charge non-conservation anomaly for both types of DWs. The main origin of this anomaly is the charge imbalance between adjacent DWs, which reminisces the charge pumping between bulk and boundaries of strained WSMs, where $\bm{B}_5$ changes sign \citep{PhysRevX.6.041021,PhysRevX.6.041046,PhysRevB.103.035306}. Therefore, this anomalous local charge density is distinct from the intrinsic charge density induced by lowest pseudo-Landau level. The sign of the anomalous charge density (electron charge or hole charge density) is directly determined by the sign of the pseudo-magnetic field. Finally, the physical continuity equation for the total charge is established by summing over all the domain walls  [SM: VII.B].
\section{Summary}
In summary, we have demonstrated that both DWs and ferromagnetic domains would undergo unusual structural changes due to electrically induced STT in magnetic WSMs. We have investigated the spatiotemporal modulation of magnetic texture and its coupling to topological itinerant electrons in the dynamical regime of DWs utilizing the quantum kinetic theory. This has resulted in the novel and non-linear anomalous Hall effect (NAHE) and non-linear anomalous drift current (ADC). NAHE and ADC can serve as a direct probe of the magnetization dynamics, while the spatial inhomogeneity of the signal can be used to probe the location of the DW. 
 %The ability to classify materials using nonlinear response is quite effective, and it also enables the detection of novel physics and dynamics of a complex system that are undetectable using linear response. 
Moreover, the NAHE, in comparison to the linear AHE which is determined by the nodes separation $|k_0|$, is proportional to $|k_0|^3$ and survives as long as the DW bears the SPT. Moreover, in contrast to the conventional drift current which is always linear in $E$, the ADC is a non-linear, local and transient signal in the dynamical regime of DWs. Therefore, our results provide an experimental pathway for exploring the domain structure by using local and non-linear probes. As a result, the real space profile of such non-linear transient signals can be used to investigate the dynamics of the magnetic texture. In a future publication, the impact of electron-electron interactions will be examined.

\section{acknowledgments}
%	\acknowledgments
SH is grateful to Dr. Azadeh Faridi for useful discussions. DC is supported by the Australian Research Council Future Fellowship FT190100062.

\appendix
	\section{Spin-Transfer Torque} \label{appa}
A non-equilibrium spin polarization of electrons $\braket{\bm{\sigma}(r)}$ leads to a chiral current in Weyl semimetals, i.e., $\bm{J}_5=ev_{\rm F} \braket{\bm{\sigma}(r)}$. Then the electrically induced STT is given by
$
\bm{T}_e=\frac{{\cal J}S}{\hbar v_{\rm F}} \hat{M} \times \braket{\bm{\sigma}(r)}=\frac{|k_0|}{e\rho_s} \hat{M} \times \bm{J}_5,
$
where $\rho_s$ is the number of local magnetic elements per unit volume. Therefore, in the absence of a magnetic field, we may write
\begin{equation} \label{dyn}
	\dfrac{d \hat{M}}{dt}=\dfrac{k_0}{\rho_s} \hat{M} \times \bm{J}_5+\alpha \hat{M} \times \dfrac{d\hat{M}}{dt}.
\end{equation}
The first term is the precession term that makes the magnetic moments rotate around an axial current $\bm{J}_5$, while the second term opposes this dynamical spin rotation with damping coefficient $\alpha$. Therefore, for the magnetization dynamics to be non-vanishing in the absence of a real magnetic field, the chiral current $\bm{J}_5$ must be generated in the system. This chiral current can be induced by an external electric field through a conventional Hall effect in the presence of an axial magnetic field $\bm{B}_5$, i.e., $\bm{J}_5=\sigma^{(H)} \hat{B}_5 \times \bm{E}$ leading to the electrically induced STT \citep{Kurebayashi_2019}
\begin{equation} \label{STT}
	\bm{T}_e=\dfrac{\sigma^{(H)}}{e \rho_s} \hat{M}\times[({\bf E}\cdot{\bf \nabla}) \hat{M} - {\nabla}(\hat{M}\cdot{\bf E})],
\end{equation}
where $\sigma^{(\text{H})}_{\text{SC}}=-\frac{v_F \tau^2 e^3 \mu}{6\pi^2 \hbar^3} |B_5|$. The full dynamical behavior of magnetic moments described in Eq. \ref{dyn} can be written in a more compact form
\begin{equation}
	\dfrac{d \hat{M}}{dt}=(1+\alpha \hat{M} \times + \alpha^2 \hat{M} \times \hat{M} \times + \cdots )\bm{T}_e= \exp(\alpha \hat{\Theta}) \bm{T}_e ,
\end{equation}
where $\hat{\Theta}=\hat{M} \times $. The above equation tells us the magnetic moment dynamics is provided by the STT of the background electrons, $\bm{T}_e$, and the other terms indicate the dissipation which opposes $\bm{T}_e$.

\begin{figure}[t]
	\includegraphics[width=0.5\textwidth]{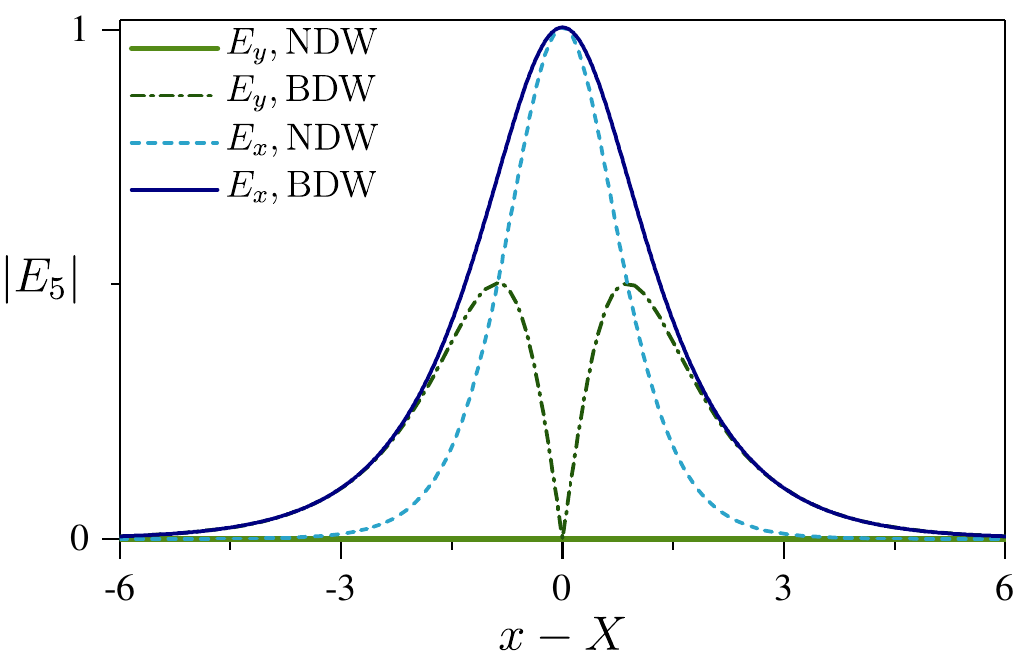}
	\caption{ The strength of the axial electric fields for 1D N\'{e}el and Bloch domain walls (NDW/BDW) in terms of the distance $x$ from the domain wall center $X$. An external electric field along the $x-$axis induces a finite $\bm{E}_5$ for both N\'{e}el and Bloch domain walls represented by solid and dashed blue lines. On the contrary,  an external electric field along the $y-$axis $\bm{E}=E_y \hat{y}$ can not induce $\bm{E}_5$ for the NDW (solid light-green line), and its value is zero at the center of the BDW (dashed dark-green line) leading to a cusp structure at $x=X$ and two maximum peaks at the opposite sides of the center.
	} \label{fig_1s}
\end{figure}

The magnetic-moment dynamics can also be understood in terms of an axial electric field $\bm{E}_5(r,t)=-\dot{\bm{A}}_5(r,t)=-\frac{{\cal J}S}{e v_{\rm F}} \dot{\hat{M}}(r,t)$ \citep{Araki_2019, Kurebayashi_2019}, which couples to the Weyl fermions with opposite sign in the same manner as an axial magnetic field $\bm{B}_5$. The axial electric field induced by magnetic dynamics is given by
\begin{equation} \label{AEF}
	\bm{E}_5=-\dfrac{\hbar |k_0|^2}{e^2 \rho_s} \sigma^{(H)} \lbrace [\hat{B}_5 + \alpha (\hat{M} \times \hat{B}_5)] (\hat{M} \cdot \bm{E}) -[\bm{E}+\alpha (\hat{M}\times \bm{E})](\hat{M}\cdot \hat{B}_5)\rbrace.
\end{equation}
It is worth noting that since the Gilbert damping rate $\alpha$ is relatively small, the effective terms are up to the linear in $\alpha$ and then we neglected the terms higher than ${\cal O}(\alpha^2)$. Accordingly, the $E_x$-induced axial electric field $\bm{E}_5$ is maximum at the domain wall center for both the N\'{e}el and Bloch domain walls [Fig. \ref{fig_1s}: solid and dashed blue lines].
On the contrary, according to this figure, applying an electric field along the $y$-axis can not induce $\bm{E}_5$ for the NDW, i.e., $\bm{E}_5^{\text{NDW}}\mid_{\text{all} \ x}^{E_y}=0$, and its magnitude tends to zero for the BDW at $x=X$, i.e., $\bm{E}_5^{\text{BDW}}\mid_{x=X}^{E_y}=0$. It means that $E_y$ is unable to induce dynamical motion to the spin moment right at the center of the BDW, and it leaves the NDW quite motionless since $E_y$ can not generate STT on the NDW, i.e., $\bm{T}_e^{\text{NDW}}\mid_{E_y}=0$.
Furthermore, the electric field along $\hat{x}$ causes a N\'{e}el domain wall to vanish entirely. To understand this, we examine the axial electric field up to linear order in $\alpha$,
\begin{equation}
	{\bf E}_5^{\text{NDW}}=-\frac{\hbar \eta}{e} \frac{|k_0|^3}{w}E_x\sech^{2}(\frac{x-X}{w})
	\hat{M}^\prime(x), 
\end{equation}
where $\hat{M}^\prime(x)=(\alpha \sech^4(\frac{x-X}{w})\sinh(\frac{x-X}{w}),1,\alpha \sech^4(\frac{x-X}{w}))$ and the time-dependent spin moment is
\begin{eqnarray}
	\hat{M}(x,t)\mid_{E_x}^{\text{NDW}}=\dfrac{\hat{M}(x)+atE_x\hat{M}^\prime(x)}{\sqrt{\sum_{i} \mid \hat{M}_i-atE_x \hat{M}^\prime \mid^2}}.
\end{eqnarray}
In the steady state, $atE_x \rightarrow \infty$, and far from the center, the second term dominates, hence $\hat{M}(x,t)\mid_{E_x}^{\text{NDW}}=(0,1,0)$. In addition, the structure arising from the components of ${\bf E}_5$ near the center is given by $\alpha(y,1/\alpha,1-2y^2)/\sqrt{1+\alpha^2}$ where $y=x-X$ tends to zero.

The presence of an external electric field $\bm{E}$ and intrinsic $\bm{B}_5$ is crucial to generate spin dynamics or axial electric field $\bm{E}_5$. It is in contrast to the strain-induced $\bm{E}_5$ in Weyl semimetals which may exist even in the absence of $\bm{B}_5$ \citep{Ilan_2019}. Furthermore, in our system the sign of $\bm{E}_5$ in space is determined by the sign of $\hat{B}_5$, i.e., $\text{sign}(E_5) \propto \text{sign}(B_5)$, then we conclude that both axial electric and magnetic fields average to zero after summing over the whole sample, i.e.,  $\sum\limits_{\text{sample}} \bm{B}_5=\sum\limits_{\text{sample}} \bm{E}_5=0$.
\section{The rate of magnetic evolution} \label{appb}
\begin{figure*}[t]
	\includegraphics[width=\textwidth]{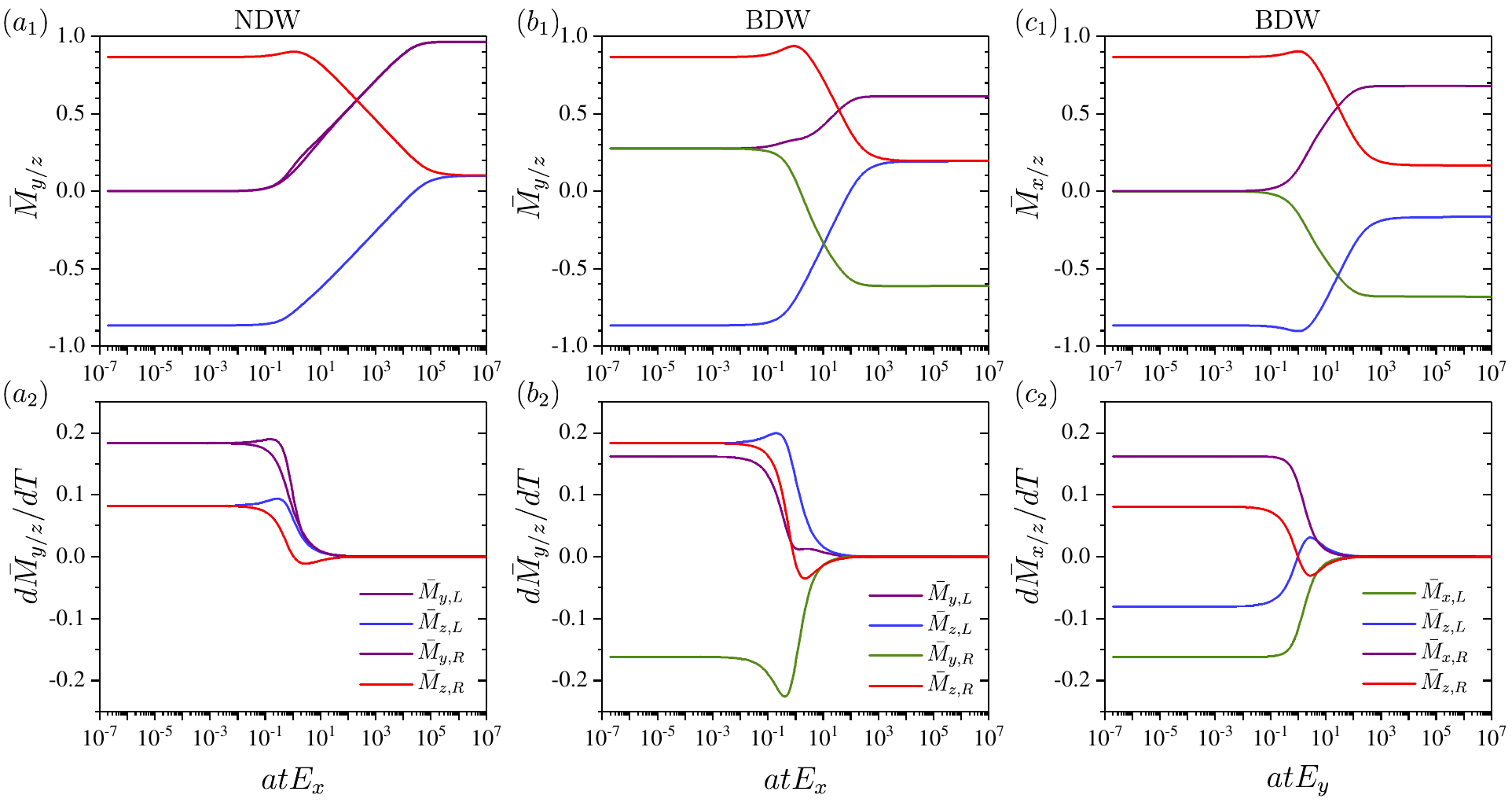}
	\caption{ ($a_1),(b_1),(c_1)$: The Log-scale representation of change in magnetic components with respect to the time multiplied to the electric field along the $x$ and $y$ directions where $a=(e \alpha \tau^2 v_F \mu |k_0|^2 )/(6 \pi^2 \rho_s \hbar^2 w)$. In the rigid motion limit, the average of the magnetic component $\bar{M}_i$ is constant, while it starts to change by increasing the parameter $aEt$ and then approach their steady-state conditions. ($a_2),(b_2),(c_2)$: The velocity of magnetic phase transition $\frac{d \bar{M}_i}{dT}$ where $T=aEt$.  The drop in $\frac{d \bar{M}_z}{dT}$ means the onset of the structural transition when $\bm{E} \parallel \hat{x}$. The label $L$ and $R$ refer to the regions $-L<x<0$ and $0<x<L$, respectively. Note that the labels BDW and NDW correspond to the initial configuration of domain walls, since both will bear the structural transition. 
	} \label{fig_2s}
\end{figure*}

As discussed in the paper, the various regimes describing the time evolution of the magnetic texture are determined by the electric field strength and its direction. For weak electric fields and short time-scales, domain walls start moving rigidly without structural deformation, which is confirmed by several experimental observations \citep{Parkin_2008,Meier_2007,Kl_ui_2005}. The numerical results in Fig. \ref{fig_2s} ($a_1,b_1,c_1$) represent the evolution of the magnetic vector components as a function of $aEt$. The parameter $\bar{M}_i$ denotes the average of the $i^{th}$ component of the magnetic vector. For the structural transition to occur one needs to exceed a threshold electric field $E_t$ and a critical time $t_c$, i.e., $tE>t_c E_t$ where $t_c \bm{E}_t=a^{-1}\frac{c_0}{c_1} \hat{x}$, $a=(e \alpha \tau^2 v_F \mu |k_0|^2 )/(6 \pi^2 \rho_s \hbar^2 w)$, $c_0= \int |\nabla_x \hat{M}|^2 \ dx$ and
$c_1= \int (|\nabla_x \hat{M}|^2-|\nabla_x \hat{M}_x|^2)\  dx$ \citep{Kurebayashi_2019}. Above this threshold value, the domain wall changes its structure and then the ferromagnetic domains with $\pm \hat{M}_z$ (at $t=0$) begin to undergo a magnetic transition to the new ferromagnetic configuration with $\hat{M}_{x,(y)}$. Fig. \ref{fig_2s} ($a_2,b_2,c_2$) presents the rate of magnetic evolution as a function of time and electric field (the parameter $T$ in figure represents the product $T=aEt$). In the rigid motion regime (induced by $E_x$ in $(a_2),(b_2)$), the magnetization components change with a constant velocity, and after the Walker critical value $aE_t t_c$ the slowdown in the evolution of the $\pm \hat{M}_z$ component can be regarded as evidence of the structural transition in the topology of the domain wall below the Curie temperature $T_c$.  The beginning of the structural transition at the onset of the reduction of the wall velocity has also been reported in several experiments \citep{Parkin_2008, Kl_ui_2005, Vanhaverbeke_2008}. The rate of magnetic domain evolution would be a non-linear function of $aEt$ and all cases approach a steady-state condition where $\frac{d \bar{M}}{d(aEt)} \rightarrow 0$. 

The time evolution is needed to reach the structural deformation in a fixed electric field; the stronger the electric field, the sooner the Walker-breakdown regime is approached and vice-versa. 

Structural transitions in the domain wall configuration have received considerable attention. The increase in domain wall mobility and width has also been attributed to increasing temperature \citep{Howlader_2020, Lee_2022, Meier_2007, Destraz_2020}. Our theoretical study shows that the domain wall dynamical regime, which includes the rigid domain wall motion and the Walker-breakdown regime, depends on the domain wall type, which was also confirmed by experimental evidence \citep{Meier_2007,Kl_ui_2005,Vanhaverbeke_2008}.
It is worth stressing that the magnetic texture evolution is plotted on a logarithmic scale as a function of $atE$ and it becomes extremely slow in the regime of large $atE$. 

\section{Contribution of magnetic textures to the kinetic theory} \label{appc}
Now, we take the effect of magnetic texture into account. If the spatial and temporal variation in $\hat{M}(\bm{r},t)$ be slow enough, the interaction of electron's spin with local magnetic texture can be taken into account through the concept of the axial vector potential $\bm{A}_5(\bm{r},t)=({\cal J}S/ev_{\rm F}) \hat{M}(\bm{r},t)$. This axial vector potential leads the momentum $\bm{p}=-i\hbar \nabla$ to transform $\bm{p}\rightarrow \bm{p}-\chi \bm{A}_5=\bm{p}-\chi ({\cal J}S/e v_{\rm F}) \hat{M}(\bm{r},t)$, where $\chi=\pm 1$ denotes chirality. 
The electric-field-induced driving term is given by
${\cal D}_{E}(\braket{\rho_i})=\frac{e\bm{E}}{\hbar} \cdot \frac{D\braket{\rho_i}}{Dk},$
and following the same method in \citep{Culcer_2017} and replacing $\chi \bm A_5$ instead of $e\bm A$, the driving force due to the domain-wall-dynamics would be
${\cal D}_{M_t}(\braket{\rho_i})=\frac{-\chi {\cal J}S}{\hbar v_{\rm F}} \dot{\hat{M}}(\bm{r},t) \cdot \frac{D\braket{\rho_i}}{Dk},$
and 
${\cal D}_{M_r}(\braket{\rho_i})=\frac{\chi {\cal J}S}{2 v_{\rm F} \hbar^2} \lbrace (\frac{D H_0}{Dk}\times (\nabla \times \hat{M}(r))\cdot \frac{D\braket{\rho_i}}{D k}\rbrace ,$
arises from the spatially inhomogeneous magnetic configuration in a ferromagnetic Weyl semimetal. Here we define the covariant derivative notation $\frac{D {\cal O}}{Dk}=\nabla_k {\cal O}-i[\bm{{\cal R}}_k,{\cal O}]$ where $\bm{{\cal R}}_k=\sum\limits_{a=x,y,z} {\cal R}_k^a e_a$ with $[{\cal R}_k^a]^{mn}=i\braket{u_k^m|\partial_{k_a}u^n_k}$.
\subsection{The matrix elements of temporal magnetic driving term: ${\cal D}_{M_t}(\braket{\rho_i})$} \label{appc1}
The matrix elements of this driving term can be written as 
\begin{equation}
	\begin{split}
		&\braket{n|{\cal D}_{M_t}(\braket{\rho_0})|n^\prime}=\dfrac{-\chi {\cal J}S}{\hbar v_{\rm F}} \bra{n} \exp(\alpha \hat{\Theta})\bm{T}_e \cdot 
		\dfrac{D \braket{\rho_i}}{D k}\ket{n^\prime}\\ &=\dfrac{-{\cal J}S}{\hbar v_{\rm F}}  (\sum_{m^\prime} \nabla_k f_{0m^\prime} \braket{n|m^\prime}\braket{m^\prime|n^\prime}+\sum_{m^\prime} f_{0m^\prime}  \bra{n}[\ket{\nabla_k m^\prime}\bra{m^\prime}+\ket{m^\prime}\bra{\nabla_k m^\prime}]\ket{n^\prime}) \cdot \exp(\alpha \hat{\Theta})\bm{T}_e .
	\end{split}
\end{equation}
Using the fact that $\nabla_k \braket{n|n^\prime}=\nabla_k \delta_{n,n^\prime}=0$, we get
\begin{equation}
	\braket{n|{\cal D}_{M_t}(\braket{\rho_0})|n^\prime}=\dfrac{-\chi {\cal J}S}{\hbar v_{\rm F}}  (\bm{\nabla}_k f_{0n} \delta_{nn^\prime}-i(f_{0n^\prime}-f_{0n}) \bm{{\cal R}}_k^{nn^\prime})\cdot \exp(\alpha \hat{\Theta}) \bm{T}_e .
\end{equation}
According to the above result, the first term represents the diagonal components of the temporal magnetic driving term and the second term denotes the off-diagonal part. The real electric field plays the same role as an axial electric field in the driving term, so we can write
\begin{equation}
	\braket{n|{\cal D}_{E}(\braket{\rho_0})|n^\prime}= e\bm{E} \cdot (\bm{\nabla}_k f_{0n} \delta_{nn^\prime}-i(f_{0n^\prime}-f_{0n}) \bm{{\cal R}}_k^{nn^\prime}) .
\end{equation}
\subsection{The matrix elements of spatial magnetic driving term: ${\cal D}_{M_r}(\braket{\rho_i})$} \label{appc2}
By making use of the magnetic-moment configuration at the domain wall as
$
\hat{M}(x)=( (1-\beta) \sech(\frac{x-X}{w}), \beta \sech(\frac{x-X}{w}), \tanh(\frac{x-X}{w}))
$, the corresponding axial magnetic field would be
\begin{equation}
	\bm{B}_5 (x)=\dfrac{{\cal J}S}{v_{\rm F}} \nabla \times \hat{M}(x)=-\dfrac{{\cal J}S}{w v_{\rm F}} \sech^2(\dfrac{x-X}{w}) (0,1,\beta \ \sinh( \dfrac{x-X}{w})
\end{equation}
which changes sign with chirality. This axial magnetic field is localized at the domain wall and near the center ($x \rightarrow X$) $\bm{B}_5$ or the curl of the magnetization lying along the y-axis.
Therefore, the spatial magnetic driving term due to the non-vanishing curl of the local magnetization at the domain wall can be written as
\begin{equation}
	{\cal D}_{M_r}(\braket{\rho_i})=\dfrac{\chi {\cal J}S}{2 v_{\rm F} \hbar^2} \lbrace (\dfrac{D H_0}{Dk}\times (\nabla \times \hat{M}(r))\cdot \dfrac{D\braket{\rho_i}}{D k}\rbrace= \dfrac{\chi |B_{5,y}|}{2 \hbar^2} [\lbrace \dfrac{DH_0}{Dk_x},\dfrac{D\braket{\rho_i}}{Dk_z}\rbrace-\lbrace \dfrac{DH_0}{Dk_z},\dfrac{D\braket{\rho_i}}{Dk_x}\rbrace].
\end{equation}
Here $\lbrace \ , \ \rbrace$ represents the matrix anticommutator. In order to calculate spatial magnetic driving term originating from the diagonal part of the density matrix $\braket{\xi}$, one can replace $\braket{\xi} \rightarrow \braket{\rho_i}$ in the above equation and then obtain its matrix elements. 
In the eigenstate basis, we have
$
\dfrac{D H_0}{D k_x}=\sum\limits_m \partial_{k_x} \epsilon_m \ket{m}\bra{m}+\sum\limits_m \epsilon_m [\ket{\partial_{k_x}m}\bra{m}+\ket{m}\bra{\partial_{k_x}m}],
$
and
$
\dfrac{D \braket{\xi}}{D k_z}=\sum\limits_{m^\prime} \partial_{k_z} f_{0m^\prime} \ket{m^\prime}\bra{m^\prime}+ \sum\limits_{m^\prime} f_{0m^\prime}  [\ket{\partial_{k_z}m^\prime}\bra{m^\prime}+ \ket{m^\prime}\bra{\partial_{k_z}m^\prime}].
$

Then we can find
	\begin{equation}
		\begin{split}
			& \lbrace \dfrac{DH_0}{Dk_x},\dfrac{D\braket{\xi}}{Dk_z}\rbrace=\dfrac{DH_0}{Dk_x} \dfrac{D\braket{\xi}}{Dk_z} +\dfrac{D\braket{\xi}}{Dk_z} \dfrac{DH_0}{Dk_x}=\\ & =\begin{pmatrix}
				2 (\partial_x \epsilon_k^n)(\partial_z \xi_k^n)+g^{nn}_{zx} & -i {\cal R}_x^{nn^\prime} (\epsilon_k^{n^\prime}-\epsilon_k^n) \dfrac{\partial}{\partial k_z} (\xi_k^n+\xi_k^{n^\prime})+g^{nn^\prime}_{zx} \\ \\
				i {\cal R}_x^{n^\prime n} (\epsilon_k^{n^\prime}-\epsilon_k^n) \dfrac{\partial}{\partial k_z} (\xi_k^n+\xi_k^{n^\prime})+g^{n^\prime n}_{zx} & 2 (\partial_z \xi_k^{n^\prime})(\partial_x \epsilon_k^{n^\prime})+g^{n^\prime n^\prime}_{zx}
			\end{pmatrix},
		\end{split}
	\end{equation}
	and 
	\begin{equation}
		\begin{split}
			& \lbrace \dfrac{DH_0}{Dk_z},\dfrac{D\braket{\xi}}{Dk_x}\rbrace=\dfrac{DH_0}{Dk_z} \dfrac{D\braket{\xi}}{Dk_x} +\dfrac{D\braket{\xi}}{Dk_x} \dfrac{DH_0}{Dk_z}=\\ & =\begin{pmatrix}
				2 (\partial_x \xi_k^n)(\partial_z \epsilon_k^n)+g^{nn}_{xz} & -i {\cal R}_z^{nn^\prime} (\epsilon_k^{n^\prime}-\epsilon_k^n) \dfrac{\partial}{\partial k_x} (\xi_k^n+\xi_k^{n^\prime})+g^{nn^\prime}_{xz} \\ \\
				i {\cal R}_z^{n^\prime n} (\epsilon_k^{n^\prime}-\epsilon_k^n) \dfrac{\partial}{\partial k_x} (\xi_k^n+\xi_k^{n^\prime})+g^{n^\prime n}_{xz} & 2 (\partial_x \xi_k^{n^\prime})(\partial_z \epsilon_k^{n^\prime})+g^{n^\prime n^\prime}_{xz}
			\end{pmatrix},
		\end{split}
	\end{equation}
	where
	\begin{equation}
		\begin{split}
			& g_{zx}^{nn}=({\cal R}_x^{nn^\prime} {\cal R}_z^{n^\prime n}+{\cal R}_x^{n^\prime n} {\cal R}_z^{n n^\prime}) (\epsilon_k^{n^\prime}-\epsilon_k^n)(\xi_k^{n^\prime}-\xi_k^n), \\
			& g_{zx}^{nn^\prime}=-i{\cal R}_z^{nn^\prime} (\xi_k^--\xi_k^+) \dfrac{\partial}{\partial_{k_x}} (\epsilon_k^{n}+\epsilon_k^{n^\prime}),\\
			& g_{zx}^{n^\prime n^\prime}=({\cal R}_x^{n^\prime n} {\cal R}_z^{n n^\prime}+{\cal R}_x^{n n^\prime} {\cal R}_z^{n^\prime n}) (\epsilon_k^{n^\prime}-\epsilon_k^n)(\xi_k^{n^\prime}-\xi_k^n), \\
			& g_{zx}^{n^\prime n}=-i{\cal R}_z^{n^\prime n} (\xi_k^--\xi_k^+) \dfrac{\partial}{\partial_{k_x}} (\epsilon_k^{n}+\epsilon_k^{n^\prime}).
		\end{split}
	\end{equation}
	Finally, we get
	\begin{equation}
		{\cal D}_{M_r}(\braket{\xi})=\dfrac{\chi |B_{5,y}|}{\hbar^2} \begin{pmatrix}
			0 & -i \epsilon_k ({\cal R}_x^{n n^\prime} \partial_z-{\cal R}_z^{n n^\prime} \partial_x ) (f_0(\epsilon_k^n)+f_0(\epsilon_k^{n^\prime})) \\ \\
			i \epsilon_k ({\cal R}_x^{n^\prime n} \partial_z-{\cal R}_z^{n^\prime n} \partial_x ) (f_0(\epsilon_k^n)+f_0(\epsilon_k^{n^\prime})) & 0
		\end{pmatrix}
	\end{equation}

or equivalently
\begin{equation} \label{Dmr}
	{\cal D}^{nn^\prime}_{M_r}(\braket{\xi})=\dfrac{-i \chi |B_{5,y}|}{\hbar^2} \epsilon_k ({\cal R}_x^{n n^\prime} \partial_z-{\cal R}_z^{n n^\prime} \partial_x ) (f_0(\epsilon_k^n)+f_0(\epsilon_k^{n^\prime})),
\end{equation}
where we have used the fact that $\epsilon_k^n=-\epsilon_k^{n^\prime}=\epsilon_k$.

\section{Density Matrix Elements: Dynamical regime} \label{appd}
Now, we assume that the external electric field makes the magnetic-moment of magnetic domain walls to be evolved in time so we have non-zero $\bm{E}_5$ as well as a non-zero external electric filed $\bm{E}$. We also assume that in the absence of $\bm{B}_5$, the diagonal and off-diagonal parts of the density matrix can be separated as $\braket{\rho_t}=\braket{\xi_t}+\braket{S_t}$, where $\braket{\rho_t}$ is the deviation of density matrix from its equilibrium condition $\braket{\rho_0}$.
The kinetic equation in this case would be
\begin{equation}
	\dfrac{\partial \braket{\rho_t}}{\partial t}+\dfrac{i}{\hbar}[H_0,\braket{\rho_t}]+\kappa(\braket{\rho_t})={\cal D}_{M_t}(\braket{\rho_0})+{\cal D}_{E}(\braket{\rho_0}).
\end{equation}
In the steady-state condition the diagonal part of the kinetic equation is written as $\kappa^{nn}_k(\braket{\xi_t})=\braket{\xi_t}^n_k/ \tau_k^n=[{\cal D}_{M_t}(\rho_0)]_k^{nn}+[{\cal D}_{E}(\rho_0)]_k^{nn}$, which leads to
\begin{equation} \label{dt1}
	\braket{\xi_t}^{nn}_k=\dfrac{\chi {\cal J}S}{\hbar v_{\rm F}} \tau_k^n (-\partial f_{0n}/\partial \epsilon_k^n) (\bm{v}_k^n \cdot \exp(\alpha \hat{\Theta}))\bm{T}_e 
	-e \tau_k^n \bm{E}\cdot \bm{v}_k^n (-\partial f_{0n}/\partial \epsilon_k^n)=\braket{\xi_{M_t}}^{nn}_k+\braket{\xi_E}^{nn}_k,
\end{equation}
where the second term originally stems from the external electric field. The off-diagonal part of the density matrix satisfies the following equation
\begin{equation}
	\dfrac{\partial \braket{S_t}^{nn^\prime}}{\partial t}+\dfrac{i}{\hbar}[H_0,\braket{S_t}]^{nn^\prime}+\kappa^{nn^\prime}_k(\braket{\xi_t})={\cal D}_{M_t}^{nn^\prime}(\braket{\rho_0})+{\cal D}_{E}^{nn^\prime}(\braket{\rho_0}),
\end{equation}
where, for simplicity, we have assumed that the off-diagonal part of the density matrix is independent of the weak disorder, i.e., $\kappa^{nn^\prime}_k(\braket{S_t})=0$, and according to the previous arguments the term ${\cal D}_{t}^{nn^\prime}(\braket{\rho_0})={\cal D}_{M_t}^{nn^\prime}(\braket{\rho_0})+{\cal D}_{E}^{nn^\prime}(\braket{\rho_0})$ for $n \neq n^\prime$ is obtained as
\begin{equation}
	{\cal D}_{t}^{nn^\prime}(\braket{\rho_0})=  i \dfrac{\chi{\cal J}S}{\hbar v_{\rm F}} (f_{0n^\prime}-f_{0n}) \bm{{\cal R}_k}^{nn^\prime} \cdot \exp(\alpha \hat{\Theta}) \bm{T}_e 
	+i \dfrac{e}{\hbar} \bm{E} \cdot \bm{{\cal R}_k}^{nn^\prime} (f_{0n^\prime}-f_{0n}),
\end{equation}
where $\bm{{\cal R}_k}^{nn^\prime}=i \braket{n|\nabla_k n^\prime}$. Following the same methods as before, we find that
\begin{equation}
	\braket{S_t}_k^{nn^\prime}=\int_0^\infty dt^\prime e^{-iH_0 t^\prime/\hbar}[{\cal D}_{t}(\braket{\rho_0})-\kappa(\braket{\xi_t})]_k^{nn^\prime}  e^{iH_0 t^\prime/\hbar}.
\end{equation}
This time integral can be easily evaluated and and the answer is provided by
$\braket{S_t}_k^{nn^\prime}=-i\hbar \dfrac{[{\cal D}_{t}(\braket{\rho_0})-\kappa(\braket{\xi_t})]_k^{nn^\prime}}{\epsilon_k^n-\epsilon_k^{n^\prime}}, $
or equivalently
\begin{equation} \label{dt2}
	\braket{S_t}_k^{nn^\prime}=-\chi k_0 \frac{f_{0n}-f_{0n^\prime}}{\epsilon_k^n-\epsilon_k^{n^\prime}} \bm{{\cal R}}_k^{nn^\prime} \cdot \exp(\alpha \hat{\theta})\bm{T}_e-e \bm{E}\cdot \bm{{\cal R}}_k^{nn^\prime} \frac{f_{0n}-f_{0n^\prime}}{\epsilon_k^n-\epsilon_k^{n^\prime}}+ i \dfrac{\kappa(\xi_E)+\kappa(\xi_{M_t})}{\epsilon_k^n-\epsilon_k^{n^\prime}}
	=\braket{S_{M_t}}_k^{nn^\prime}+\braket{S_E}_k^{nn^\prime},
\end{equation}
where $\braket{S_{M_t}}_k^{nn^\prime}$ and $\braket{S_E}_k^{nn^\prime}$, respectively, take into account the spin-dynamic-induced and external electric-field-induced off-diagonal density matrices.
\section{Inter-band current and non-linear anomalous Hall effect} \label{appe}
In order to calculate the current, we first define the velocity and density matrix tensor in the eigen-states basis ($+$ for conduction and $-$ for valance bands) as the following
\begin{equation}
	\tensor{\bm{V}}_k=\begin{pmatrix}
		\hbar^{-1} \nabla_k \epsilon_k^+ \ \ & \ \ i\hbar^{-1} (\epsilon_k^+-\epsilon_k^{-}) \bm{{\cal R}}_k^{+ -}\\ \\
		-i \hbar^{-1} \bm{{\cal R}}_k^{- +}(\epsilon_k^+-\epsilon_k^{-}) \ \ & \ \ \hbar^{-1} \nabla_k \epsilon_k^{-}
	\end{pmatrix}
	=\begin{pmatrix}
		\bm{V}_k^{++} \ \ & \ \ \bm{V}_k^{+-} \\ \\
		\bm{V}_k^{-+} \ \ & \ \ \bm{V}_k^{--}
	\end{pmatrix},
\end{equation}
and
$
\tensor{\braket{\rho}}_k=\begin{pmatrix}
	\braket{\xi}^{++}_k \ \ & \ \ \braket{S}^{+-}_k \\ \\
	\braket{S}^{-+}_k \ \ & \ \ \braket{\xi}^{--}_k
\end{pmatrix},
$
where $\braket{\xi}^{nn}_k$ and $\braket{S}^{nn^\prime}_k$ are the diagonal and off-diagonal components of the density matrix. Then we can obtain
\begin{equation}
	\tensor{\bm{j}_k}= \tensor{\bm{V}}_k \tensor{\braket{\rho}}_k =\begin{pmatrix}
		\bm{v}_k^{+} \braket{\xi}^{++}_k+\bm{V}_k^{+-} \braket{S}_k^{-+} \ \ & \ \ \bm{v}_k^{+} \braket{S}_k^{+-}+\bm{V}_k^{+-} \braket{\xi}_k^{--} \\ \\
		\bm{V}_k^{-+} \braket{\xi}_k^{++}+\bm{v}_k^{-} \braket{S}_k^{-+}  \ \ & \ \
		\bm{V}_k^{-+} \braket{S}_k^{+-}+\bm{v}_k^{-} \braket{\xi}_k^{--}
	\end{pmatrix}=\begin{pmatrix}
		\bm{j}^{++}_k  \ \ & \ \ \bm{j}_k^{+-} \\ \\ 
		\bm{j}_k^{-+} \ \ & \ \ \bm{j}_k^{--}
	\end{pmatrix},
\end{equation}
where we have used $\bm{V}_k^{nn}=\bm{v}_k^{n}$.
The current density as an observable quantity is exactly the trace of the above matrix, since we have $\bm{J}=\text{Tr}[\tensor{\bm{j}_k}]= \bm{j}_k^{++}+\bm{j}_k^{--}=\sum_n \int [d\bm{k}] \bm{j}_k^{nn},$
where Tr$=\sum\limits_n \int [d\bm{k}]$.
According to the above definition, the current density associated to node $\chi$ is given by
\begin{equation}
	\bm{J}_\chi= 
	\text{Tr}[-e\bm{j}^{nn}_{\chi,k}]=-e \sum_n \int [dk] \bm{v}_k^n \braket{\xi}^{nn}_{\chi,k}-e \sum_{n,n^\prime} \int [dk] \bm{V}_k^{nn^\prime} \braket{S}^{n^\prime n}_{\chi,k}.
\end{equation}

The inter-band contribution to the current comes from the band off-diagonal terms in the density matrix $\braket{S_t}_k^{nn^\prime}=\braket{S_E}_k^{nn^\prime}+\braket{S_{M_t}}_k^{nn^\prime}$ [Eq. \ref{dt2}] in response to an external electric field (labeled by subscript $E$) and magnetic texture dynamics (labeled by subscript $t$) is given by
\begin{equation} \label{ddin}
	\braket{S_t}_k^{nn^\prime}=-\chi k_0 \frac{f_{0n}-f_{0n^\prime}}{\epsilon_k^n-\epsilon_k^{n^\prime}} \bm{{\cal R}}_k^{nn^\prime} \cdot \exp(\alpha \hat{\theta})\bm{T}_e-e \bm{E}\cdot \bm{{\cal R}}_k^{nn^\prime} \frac{f_{0n}-f_{0n^\prime}}{\epsilon_k^n-\epsilon_k^{n^\prime}}+ i \dfrac{\kappa(\xi_E)+\kappa(\xi_t)}{\epsilon_k^n-\epsilon_k^{n^\prime}}.
\end{equation}
Having integrated the current $\bm{J}^{\text{IB}}=-ie \sum_{n,n^\prime,\chi}\int [dk] (\epsilon_{k,\chi}^n-\epsilon_{k,\chi}^{n^\prime}) \bm{{\cal R}}^{nn^\prime}_{\chi,k} \braket{S_t}_{\chi,k}^{n^\prime n}$, leads to the following inter-band current
\begin{equation} \label{inter}
	\bm{J}^{\text{IB}}=-\dfrac{e^2}{\hbar}|k_0|\dfrac{\hat{M}(x,t)\times \bm{E}}{f_E(x,t)}+\dfrac{e^2}{\hbar} \int_0^t \dfrac{\bm{E}_5(x,t^\prime) \times \bm{E}}{f_E(x,t)}dt^\prime- \dfrac{e^2}{\hbar} \sum_{n,n^\prime,\chi} \int [dk] \bm{{\cal R}}_k^{n n^\prime} (\kappa^{n^\prime n}_k(\braket{\xi_{M_t}})+\kappa^{n^\prime n}_k (\braket{\xi_E})),
\end{equation}
where $f_E(x,t)=\sqrt{\sum_{i=x,y,z} \mid \hat{M}_i-e/\hbar \int_0^t \frac{\bm{E}_5(x,t^\prime)}{k_0}dt^\prime\mid ^2}$ and we have used $\epsilon_{k,\chi}^{n=\pm}=\pm \hbar v_{\rm F} |\bm{k}-\chi \bm{k}_w|$. It is easy to show that the first two terms in Eq. \ref{inter} is exactly the anomalous Hall effect in WSMs, i.e., $\bm{J}\propto \hat{M}_E(x,t) \times \bm{E}$, where $\hat{M}_E(x,t)$ is a function of $\bm E$. Then, the first and second terms represent the linear and non-linear anomalous Hall effect (we note that $|\bm{E}_5| \propto |\bm{E}|$ (Eq. \ref{AEF})), and the third term stems from the role of scattering process in the current parallel to the Berry vector potential $ \bm{{\cal R}}_k^{n n^\prime}$.
The $\bm{E}_5$-induced off-diagonal density matrix, i.e., the first term of Eq. \ref{ddin}, does not contribute to the inter-band current, i.e., an anomalous Hall effect cannot be generated by an axial electric field.

\section{Local chiral/charge non-conservation by domain wall motion} \label{appf}
Having employed the quantum kinetic theory, we show that the STT at the domain wall can activate the localized axial anomaly in magnetic Weyl semimetals with domain walls. This axial anomaly generated by $\bm{E}_5 \cdot \bm{B}_5 \neq 0$ causes an axial density $n_5$ to build up at the domain wall. The rate of electric-field-induced charge accumulation at each node is given by
\begin{equation} \label{ca1}
	\dfrac{\partial N_\chi}{\partial t}=\text{Tr}[{\cal D}_{M_r} (\braket{\rho_t})]+\text{Tr}[{\cal D}_{M_t} (\braket{\rho_r})]+\text{Tr}[{\cal D}_{E} (\braket{\rho_r})]+\text{Tr}[{\cal D}_{M_r} (\braket{\rho_E})].
\end{equation}
The first two terms are proportional to $\bm{E}_5 \cdot \bm{B}_5$ and the last two terms are proportional to $\bm{E} \cdot \bm{B}_5$. These terms are the source of creation or annihilation of charges in each valley.

\subsection{Electric-field-induced chiral anomaly: $\partial_\mu J^\mu_5\neq 0$} \label{appf1}
According to Eq. \ref{Dmr}, the diagonal driving term ${\cal D}_{M_r}^{nn}[\braket{\xi_t}]$ is zero. Then for electric field parallel to x-axis we have
\begin{equation} 
	{\cal D}_{M_r}^{nn}[\braket{S_{M_t}}]=\dfrac{\chi |B_{5,y}|}{\hbar} (v_x \partial_z-v_z \partial_x) \braket{S_t}=|k_0|^2 (\nabla \times \hat{M}) (\alpha \hat{M}\times \bm{T}_e)\cdot (v_x^n \Omega_{\chi,x}^n,0,v_z^n \Omega_{\chi,z}^n) (-\dfrac{\partial f_0(\epsilon_k)}{\partial \epsilon_k}),
\end{equation}
We note that ${\cal D}_{M_r}^{nn}[\braket{S_{E}}]=0$ for $\bm{E}\parallel \hat{x}$.  Then we get
\begin{equation}\label{caa1}
	\text{Tr}[{\cal D}_{M_r}^{nn}[\braket{S_{M_t}}]=\dfrac{|K_0|^2}{4\pi^2} (\nabla \times \hat{M})_y (\alpha \hat{M}\times \bm{T}_e)_y \sum_n \int \dfrac{d^3k}{2\pi} \delta(\mu-\epsilon_k) [v_x^n \Omega_{\chi,x}^n+v_z^n \Omega_{\chi,z}^n].
\end{equation}
The term Tr$[{\cal D}_{M_t} (\braket{\rho_r})]$ can be obtained in terms of both the diagonal and off-diagonal density matrix, i.e., Tr$[{\cal D}_{M_t} (\braket{\rho_r})]$=Tr$[{\cal D}_{M_t} (\braket{\xi_r})]+$Tr$[{\cal D}_{M_t} (\braket{S_r})]$. The only non-zero contribution comes from ${\cal D}_{M_t} (\braket{\xi_r})$ and again  we note that ${\cal D}_{E} (\braket{\xi_r})=0$ for $\bm{E} \parallel \hat{x}$. The diagonal part would be
\begin{equation}
	[{\cal D}_{M_t} (\braket{\xi_r})]^{nn}=-\chi |k_0| \exp(\alpha \hat{\Theta}) \bm{T}_e \cdot \nabla_k \braket{\xi_r}=|k_0|^2 \ (\alpha \hat{M} \times \bm{T}_e) \cdot \bm{v}_k \ (\nabla \times \hat{M})\cdot \bm{\Omega}_{\chi,k}^n \ (-\dfrac{\partial f_{0n}}{\partial \epsilon_k}),
\end{equation}
where we have used $\braket{\xi_r}=\chi |k_0| f_{0n} (\nabla \times \hat{M})\cdot \bm{\Omega}_{\chi,k}^n$. Then we have
\begin{equation} \label{ca22}
	\text{Tr}[{\cal D}_{M_t}^{nn}[\braket{\xi_r}]=\dfrac{|k_0|^2}{4\pi^2} (\nabla \times \hat{M})_y (\alpha \hat{M}\times \bm{T}_e)_y  \sum_n \int \dfrac{d^3k}{2\pi} \delta(\mu-\epsilon_k) [v_y^n \Omega_{\chi,y}^n].
\end{equation}
Combining Eq. \ref{caa1} and \ref{ca22}, we get the chiral anomaly expression generated by the domain wall motion. The rate of charge pumping between valleys would be
\begin{equation} \label{anm}
	\dfrac{\partial N_\chi}{\partial t}=\text{Tr}[{\cal D}_{M_r} (\braket{\rho_t})]+\text{Tr}[{\cal D}_{M_t} (\braket{\rho_r})]=\dfrac{|k_0|^2}{4\pi^2 \hbar^2} (\nabla \times \hat{M})\cdot (\alpha \hat{M}\times \bm{T}_e) \sum_n \int \dfrac{d^3k}{2\pi} \delta(\mu-\epsilon_k) \ \bm{v}_k^n \cdot \bm{\Omega}_{\chi,k}^n,
\end{equation}
or in a more compact form we have
\begin{equation}\label{CAN1}
	\dfrac{\partial N_\chi}{\partial t}=\dfrac{\chi}{2\pi^2 \hbar^2} \bm{E}_5(x,t) \cdot (\nabla \times \bm{M}(x,t))-\dfrac{N_\chi}{\tau_2}.
\end{equation}
The second term takes into account inter-valley scattering processes, where $\tau_2$ is the inter-valley relaxation time related to processes that can change the valley quasiparticle number, while the intra-valley process denoted by $\tau_1^{-1}$ preserves the particle number.
As expected from the chiral anomaly, the rate of change of the particle number in each valley is proportional to the chirality $\chi$ meaning that chiral fermions are created in one node and annihilated in another one.
In conclusion, in the case of $\bm{E} \parallel \hat{x}$ and using the explicit expression for axial magnetic and electric fields for the N\'{e}el and Bloch domain walls we find the field-induced anomaly as
\begin{equation} \label{CAN2}
	\dfrac{\partial N_\chi}{\partial t}\mid^{\text{NDW}}_{E_x}\simeq\chi \dfrac{\eta |k_0|^4}{2\pi^2 \hbar^2 w^2} E_x \dfrac{\sech^4(\frac{x-X}{w})}{f_E(x,t)},
\end{equation}
and $\label{CAB1}
\dfrac{\partial N_\chi}{\partial t}\mid^{\text{BDW}}_{E_x}=0,$
where $\eta=e \alpha \tau^2 v_F \mu /6 \pi^2 \rho_s \hbar^2$. 

In the long-time limit $t\gg \tau_2$, the steady state solution for each valley reads $\delta n^{(1)}_\chi=\frac{\chi \tau_2}{2\pi^2 \hbar^2} \bm{E}_5 \cdot \bm{B}_5$. In the opposite limit $t\ll \tau_2$ where the time evolution of the magnetic texture is faster than the inter-valley scattering, the particle number density at each valley is given by $\delta n^{(1)}(x,t)=\int dt^\prime \frac{\chi}{2 \pi^2 \hbar^2} \bm{E}_5\cdot \bm{B}_5(x,t^\prime)$ leading to a shift in the chemical potential
$\delta \mu^{(1)}_\chi(x,t)\simeq \dfrac{2\pi^2 \hbar^3 v_{\rm F}^3}{\mu_0^2} \delta n^{(1)}_\chi(x,t),$
where $\mu_0$ is the initial location of the chemical potential and $v_{\rm F}$ is the Fermi velocity. The STT-induced axial chemical potential, $\delta \mu_5=\delta \mu_+ - \delta \mu_-$, at the domain walls decays to zero when the domain wall structures reach their new steady-state configuration.
The axial anomaly is activated for any non-zero electric field strength, while it is strongly dependent on the field direction. Interestingly we conclude the axial anomaly generation across the NDW is an anisotropic phenomenon and only exists when $\bm{E} \parallel \hat{x}$. The axial anomaly can be also activated in magnetic Weyl semimetals by a magnetic field $\bm{B}$ beyond the Walker-breakdown field, i.e., $|B|>|B_t|$, where the internal angle of the spin moments oscillates in time leading to a structural deformation of domain wall \citep{Hannukainen_2020}.

\subsection{Electric-field-induced local charge non-conservation: $\partial_\mu J^\mu\neq 0$} \label{appf2}

For an electric field along the $y$-axis, the non-zero terms on the right-hand site of Eq. \ref{ca1} is
\begin{equation} \label{a1}
	\text{Tr}[{\cal D}_E^{nn}(\braket{\xi_r})]=\dfrac{eE_y}{\hbar} B_{5,y} \sum_n \int [dk] \delta(\mu-\epsilon) [v_y^n \Omega_{\chi,y}^n],
\end{equation}
and 
\begin{equation} \label{a2}
	\text{Tr}[{\cal D}_{M_r}^{nn}(\braket{S_E})]=e\chi |k_0| (\nabla \times \hat{M})_y E_y \sum_n \int [dk] \delta(\mu-\epsilon) [v_x^n \Omega_{\chi,x}^n+v_z^n \Omega_{\chi,z}^n].
\end{equation}
Then the rate of particle number change in each node when $\bm{E} \parallel \hat{y}$ is given by
\begin{equation} \label{a3}
	\dfrac{\partial N_\chi}{\partial t}=\dfrac{e}{2 \pi^2} \bm{E}\cdot \bm{B}_5 (x,t)-\dfrac{N_\chi}{\tau},
\end{equation}
where we denote the relevant scattering time by $\tau$.
In contrast to Eq. \ref{CAN1}, the rate of particle number change in each valley is independent of the chirality. Therefore, applying an external electric field along the $y$-axis induces the charge density 
\begin{equation} \label{n}
	\delta n^{(2)}_\chi(x,t)=\frac{e}{2\pi^2 \hbar^2} \int dt^\prime \ \bm{E} \cdot \bm{B}_5(x,t^\prime),
\end{equation}
in both nodes with the same sign, leads to a slight shift in the chemical potential
$
\delta \mu^{(2)}(x,t)=\frac{2\pi^2 \hbar^3 v_{\rm F}^3}{\mu_0^2} \int dt^\prime \ \bm{E}\cdot \bm{B}_5(x,t^\prime),
$
in the limit of $t\ll \tau$. In the steady-state condition $t\gg \tau$, on the other hand, the time-dependent function in the integral behaves adiabatically, and then it can be pulled out of the integral. For a NDW, we have $\bm{E}_5\mid^{\text{NDW}}_{E_y}=0$ and $f(x,t)\mid_{E_y}^{\text{NDW}}=1$, therefore the $E_y$-induced charge density in NDW is $\delta n^{(2)}|^{\text{NDW}}=\frac{e \tau}{2 \pi^2 \hbar^2} \bm{E}\cdot \bm{B}_5 (x,0)$. For BDW, the charge density $\delta n^{(2)}(x,t)$ can be obtained from Eq. \ref{n}.
\begin{figure}[t]
	\includegraphics[width=0.5\textwidth]{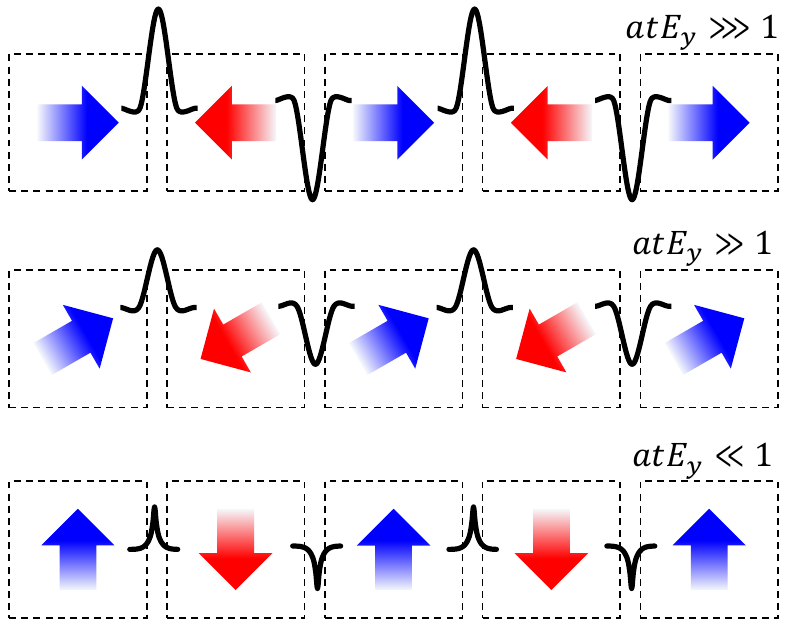}
	\caption{ The evolution of $E_y$-induced charge and discharge. The lowest panel demonstrates $\delta n^{(2)}$ localized at the BDW between domains with $\pm \hat{M}_z$ ($atE_y\ll 1$) and the highest panel is the head-to-head (or tail-to-tail) walls between domains with $\pm \hat{M}_x$ ($atE_y\ggg 1$).
	} \label{fig_3s}
\end{figure}
Since the induced charge density $\delta n_\chi^{(2)}$ couples to each node with the same sign, both nodes receive the same amount of charge due to an external electric field along the $y$-axis. This term can be interpreted as the anomalous non-conservation of the local charge at a single domain wall. The violation of local charge conservation in Weyl semimetals has also been studied in strained Weyl semimetals, which leads to the propagation of novel plasmon modes \citep{PhysRevLett.118.127601,PhysRevB.103.035306} or even the modification of phonon modes \citep{Heidari_2019}. Since the domain-wall-induced pseudo-magnetic field will vanish after summing over the entire system, the physical issue of non-conservation of charges at a single domain wall can also be solved by taking into account the contribution of all domain walls.  Figure. \ref{fig_3s} represents a simple schematic of the three snapshots of the domain wall (start from BDW at $t=0$): (i) $aEt\ll 1$, (ii) $aEt\gg 1$, and (iii) $aEt\ggg 1$ (steady-state limit). Following the general statement that the pseudo-fields should average to zero over the whole sample, the charge density $\delta n_\chi^{(2)}$ in the simple model sketched in Fig. \ref{fig_3s} must come in pairs with opposite signs.
While the charge density $\delta n^{(2)}_\chi$ is non-linear in the electric field for the BDW, it increases linearly for the NDW. 

\bibliography{ref}

\end{document}